\documentclass[aps,prd,superscriptaddress,onecolumn,floatfix,nofootinbib,preprintnumbers]{revtex4}
\pdfoutput=1

\usepackage{amsmath,graphicx,tabularx,url}

\begin{document}

\def\tablename{Table}
\def\figurename{Figure}

\def\gtot{\Gamma_\text{tot}}
\def\brinv{\text{BR}_\text{inv}}
\def\brsm{\text{BR}_\text{SM}}
\def\bratio{\mathcal{B}_\text{inv}}
\def\as{\alpha_s}
\def\az{\alpha_0}
\def\gz{g_0}
\def\w{\vec{w}}
\def\sdag{\Sigma^{\dag}}
\def\s{\Sigma}
\newcommand{\psib}{\overline{\psi}}
\newcommand{\Psib}{\overline{\Psi}}
\newcommand\one{\leavevmode\hbox{\small1\normalsize\kern-.33em1}}
\newcommand{\Mpl}{M_\mathrm{Pl}}
\newcommand{\p}{\partial}
\newcommand{\lag}{\mathcal{L}}
\newcommand{\qqquad}{\qquad \qquad}
\newcommand{\qqqquad}{\qquad \qquad \qquad}

\newcommand{\qb}{\bar{q}}
\newcommand{\matx}{|\mathcal{M}|^2}
\newcommand{\really}{\stackrel{!}{=}}
\newcommand{\msbar}{\overline{\text{MS}}}
\newcommand{\qns}{f_q^\text{NS}}
\newcommand{\lqcd}{\Lambda_\text{QCD}}
\newcommand{\met}{\slashchar{p}_T}
\newcommand{\pmiss}{\slashchar{\vec{p}}_T}

\newcommand{\st}[1]{\tilde{t}_{#1}}
\newcommand{\stb}[1]{\tilde{t}_{#1}^*}
\newcommand{\nz}[1]{\tilde{\chi}_{#1}^0}
\newcommand{\cp}[1]{\tilde{\chi}_{#1}^+}
\newcommand{\cm}[1]{\tilde{\chi}_{#1}^-}

\providecommand{\mg}{m_{\tilde{g}}}
\providecommand{\mst}{m_{\tilde{t}}}
\newcommand{\msn}[1]{m_{\tilde{\nu}_{#1}}}
\newcommand{\mch}[1]{m_{\tilde{\chi}^+_{#1}}}
\newcommand{\mne}[1]{m_{\tilde{\chi}^0_{#1}}}
\newcommand{\msb}[1]{m_{\tilde{b}_{#1}}}

\newcommand{\mev}{{\ensuremath\rm MeV}}
\newcommand{\gev}{{\ensuremath\rm GeV}}
\newcommand{\tev}{{\ensuremath\rm TeV}}
\newcommand{\fb}{{\ensuremath\rm fb}}
\newcommand{\ab}{{\ensuremath\rm ab}}
\newcommand{\pb}{{\ensuremath\rm pb}}
\newcommand{\sign}{{\ensuremath\rm sign}}
\newcommand{\ifb}{{\ensuremath\rm fb^{-1}}}

\def\slashchar#1{\setbox0=\hbox{$#1$}           
   \dimen0=\wd0                                 
   \setbox1=\hbox{/} \dimen1=\wd1               
   \ifdim\dimen0>\dimen1                        
      \rlap{\hbox to \dimen0{\hfil/\hfil}}      
      #1                                        
   \else                                        
      \rlap{\hbox to \dimen1{\hfil$#1$\hfil}}   
      /                                         
   \fi}
\newcommand{\dslash}{\slashchar{\partial}}
\newcommand{\Dslash}{\slashchar{D}}

\def\eg{{\sl e.g.} \,}
\def\ie{{\sl i.e.} \,}
\def\etal{{\sl et al} \,}

\title{Stop Reconstruction with Tagged Tops}

\author{Tilman Plehn}
\affiliation{Institut f\"ur Theoretische Physik, 
              Universit\"at Heidelberg, Germany}

\author{Michael Spannowsky}
\affiliation{Department of Physics and Institute of Theoretical Science, 
             University of Oregon, Eugene, USA}

\author{Michihisa Takeuchi}
\affiliation{Institut f\"ur Theoretische Physik, 
             Universit\"at Heidelberg, Germany}

\author{Dirk Zerwas}
\affiliation{LAL, IN2P3/CNRS, Orsay, France}

\date{\today}

\begin{abstract}
 At the LHC combinatorics make it unlikely that we will be able to
 observe stop pair production with a decay to a semi-leptonic top pair
 and missing energy for generic supersymmetric mass spectra. Using a
 Standard-Model top tagger on fully hadronic top decays we can not
 only extract the stop signal but also measure the top momentum. To
 illustrate the promise of tagging tops with moderate boost we include
 a detailed discussion of our HEPTopTagger algorithm.
\end{abstract}

\maketitle

\section{Introduction}
\label{sec:intro}

Searches for top squarks at hadron colliders aim at a fundamental
questions of electroweak symmetry breaking --- if the Higgs boson
should be a fundamental scalar, how can its mass be stabilized? In
particular, is the Higgs mass protected by some symmetry? Such
symmetries typically predict the existence of a top partner, like in
supersymmetric or little Higgs models~\cite{review,meade}. In such a
case, studying the properties of top partners allows us to unravel the
nature of such an underlying fundamental symmetry protecting the
fundamental Higgs mass at the scale of electroweak symmetry
breaking.\medskip

At the Tevatron, low-mass stop searches look for loop-induced stop
decays~\cite{stop_decays} to charm quarks and the lightest
neutralino~\cite{stops_tevatron_c}. Increasing the stop mass makes it
more promising to look for decays to a bottom jet and the lightest
chargino~\cite{stops_tevatron_b}, a final state irreducible from a
leptonic top decay. Finally, if the stop becomes heavier and the
strong decay in a top quark and a gluino is not yet kinematically
allowed, the stop can decay into a top quark and the lightest
neutralino~\cite{stop_decays}. This final state has the advantage that
at least hadronic top quarks we might be able to fully reconstruct,
which puts us into a promising position to study angular correlations
in the stop pair final state. Fully hadronic top pairs from stop
production are studied in the CMS TDR~\cite{cms_tdr}, Section~13.12,
but with the requirement of an additional lepton pair from the stop
decays. Including this lepton essentially removes all QCD
backgrounds. In this analysis we will show that such a lepton is not
needed once we apply an efficient identification of boosted tops.

There have been several suggestions as to what we might be able to say
about the nature of the stop based on a momentum reconstruction of its
visible decay products~\cite{stops_measure,weiler}; however, to date
there exists no experimentally confirmed analysis which extracts
hadronic or semi-leptonic top pairs plus missing energy at the LHC.
This means that without a viable discovery channel all of those
suggestions are bound to end up pure fiction in the era of actual LHC
data. In this paper we will first convince ourselves that in spite of
claims to the contrary there is no reason to assume that stop decays
to semi-leptonic top quarks plus missing energy will be discovered at
the LHC --- in line with the state of the art of experimental
simulations. We will then study the reach of
fat-jet~\cite{fatjet_vh,david_e,fatjet_tth} searches for purely
hadronic stop decays and their potential when it comes to
reconstructing for example the top momenta. In the Appendix we will
give a long-overdue study of a hadronic top tagger based on the
Cambridge/Aachen jet algorithm and a mass drop criterion. This
HEPTopTagger (Heidelberg--Eugene--Paris) is designed to cover
moderately boosted top quarks, as we also expect them for Standard
Model processes at the LHC~\cite{fatjet_tth}.~\footnote{The
  HEPTopTagger source code will be available from
  \url{www.thphys.uni-heidelberg.de/~plehn/heptoptagger}}

\section{Standard Semi-Leptonic Analysis}
\label{sec:old}

Using semi-leptonic top decays to extract the signature
\begin{equation}
 pp \to \st1 \st1^*
    \to ( t \nz1) \, (\bar{t} \nz1) 
    \to ( b \ell^+ \nu \nz1) \, (\bar{b} j j \nz1) 
       +( b j j \nz1) \, (\bar{b} \ell^- \bar{\nu} \nz1) 
\end{equation}
including four jets and missing energy from the irreducible top pair
production requires a detailed analysis of the two-dimensional missing
energy vector and its correlation with the visible momenta in the
final state. The stop mass we assume to be 340~GeV~\cite{weiler},
decaying with essentially 100\% branching ratio to a top quark and a
98~GeV lightest neutralino. The leading-order production rate for stop
pairs according to {\sc Pythia} is around 3.2~pb, the next-to-leading
order rate from {\sc Prospino} is 5.1~pb~\cite{prospino}. To compare
our result to the original analysis, in this section we do not apply
the NLO corrections, \ie a flat $K$ factor of 1.59.  For the same
reason we normalize our top-pair sample to 550~pb instead of the
approximate NNLO rate around 918~pb~\cite{top_rate}, corresponding to
$K=1.67$. The original semi-leptonic analysis starts from a set of
acceptance cuts requiring exactly four jets and a charged
lepton~\cite{weiler}:
\begin{alignat}{9}
p_{T, j} &>25~\gev \qqqquad
&|\eta_j| &< 4.0 \qqqquad
&\Delta R_{jj} &> 0.4 \notag \\
p_{T, \ell} &> 20~\gev 
&|\eta_\ell| &< 2.5 
&\Delta R_{j\ell} &> 0.4 \; .
\end{alignat}
The top-pair and $W$+jets backgrounds can be reduced by an additional
set of cuts, largely inspired by the usual semi-leptonic top analyses
at the Tevatron. One of the four jets should be $b$-tagged, with the
appropriate efficiency of $60\%$. The different jets, the lepton and
the missing energy vector have to be separated according
to~\cite{weiler}
\begin{alignat}{5}
\text{min}_j \Delta R_{\ell j} & < 1.5 \qqqquad
&\met &> 125~\gev \notag \\
\cos\phi( p_{T,\ell},\pmiss ) & > 0.7 
&0.8 &< \text{min}_x \Delta \phi(\pmiss, x) < 1.3 \quad (x = \ell,j)  \; ,
\end{alignat}
and the two reconstructed top decays have to be fulfilled~\cite{weiler}
\begin{alignat}{5}
|m_t^\text{rec} - m_t | &< 5~\gev \qquad
&&\text{(hadronic top)} \notag \\
|m_W^\text{rec} - m_W | &> 40~\gev \qquad
&&\text{(leptonic top veto with $m_t$ constraint for $\vec{p}_{\nu L}$~\cite{top_rec_semilep})}  \; .
\end{alignat}
The first condition identifies the hadronically decaying top while the
second condition makes sure that once we include the entire missing
energy from the leptonic top decay and the pair of neutralinos the
mass of the reconstructed top candidate does not match the physical
top mass~\cite{weiler}. It is possible to improve the leptonic top
veto for example by solving the kinematical constraints for the top
mass and requiring that this complex solution have the correct real
part as well as a vanishing imaginary
part~\cite{top_rec_semilep}. However, these details should not affect
the final outcome of our analysis, as we will see from the
discussion. In the following, this analysis setup we refer to as
`PW'~\cite{weiler}.\medskip

We simulate signal and background using {\sc Herwig}~\cite{herwig},
{\sc Pythia}~\cite{pythia} and {\sc Alpgen-Pythia}~\cite{alpgen}
including initial and final state radiation, hadronization and
underlying event. The top and stop samples we generate inclusively
without restricted decays.  For the fast detector simulation we rely
on {\sc Acerdet}~\cite{acerdet}, a reasonably reliable fast simulation
of LHC detectors which should agree well with full detector simulation
for the analysis presented here~\cite{csc_notes}. The final results
including the three leading backgrounds we show in
Table~\ref{tab:semilep}, labelled `PW'~\cite{weiler}. Compared to the
original work in Ref.~\cite{weiler} we see that the signal efficiency
is considerable lower, which is largely due to combinatorics in the
reconstruction of the hadronic top and subsequent reconstruction
hypotheses. In the next section, this will serve as the motivation to
instead use a top tagger, which we know is best suited to
automatically resolve combinatorial issues~\cite{fatjet_tth}.\medskip

Given our results we can slightly optimize the original semi-leptonic
analysis: Instead of exactly four jets, we require a minimum of four
jets to allow for example for initial state radiation.  The $b$ tag we
apply to jets with $|\eta_j| < 2.5$. Finally, the hadronic mass
reconstruction is considered successful if the three-jet invariant
mass is within 15~GeV of the nominal top quark mass, instead of
5~GeV. Again, the results are shown in Table~\ref{tab:semilep},
labelled `PSTZ'. For large stop masses we could consider applying a
significantly stiffer cut on missing energy, but as we will discuss in
the next section such a cut will leave us with essentially unknown
detector fake rates.\medskip

\begin{table}
\begin{tabular}{l|r|rrr|r|r||r}
\hline  
& $\sigma$~[pb] & $N_\text{simulated}$    
& $\epsilon_\text{PW}$                 
& $\epsilon_\text{PSTZ}$               
& $\sigma\cdot\epsilon_\text{PW}$~[fb] 
& $\sigma\cdot\epsilon_\text{PSTZ}$~[fb] 
& Ref.~\cite{weiler} \\
\hline
$\st1 \st1^*$  & 3.2 & 120000 & $(1.5\pm 0.1)\cdot 10^{-3}$  & $(1.2\pm0.03)\cdot 10^{-2}$     & 4.8  &  38   & 56 \\
\hline
$t\bar{t}$ &   550  & 500000 & $(8.6\pm 1.3)\cdot 10^{-5}$  & $(4.3\pm0.3)\cdot 10^{-4}$      & 47.3 & 237   & 20 \\
$W+4j$     &   56.5 & 397698 & $(3.5\pm 0.9)\cdot 10^{-5}$  & $(3.8\pm 0.3)\cdot 10^{-4}$     &  2.0 &  21.5 & $\sim 2.7$ \\
$W+bbjj$  &   0.63 & 761937 & $(3.1\pm 0.2)\cdot 10^{-4}$   & $(2.7 \pm 0.06)\cdot 10^{-3}$  &  0.2 &   1.7 & $\sim 1.5$ \\
\hline
SM total   &        &        &                              &                                & 49.5 & 260.2 & $\sim 24.2$ \\
\hline     
$S/B$                  &&&&& 0.096 & 0.15 & 2.3  \\
$S/\sqrt{B}_{10~\ifb}$ &&&&& 2.2  & 7.5  & 36 \\
\end{tabular}
\caption{Signal and backgrounds for the semi-leptonic stop
  analysis. The three sets of results correspond to the analysis
  suggested in Ref.~\cite{weiler} including ISR/FSR, hadronization and
  fast detector simulation (PW), a slightly modified version of the
  same analysis including ISR/FSR, hadronization and fast detector
  simulation (PSTZ), and the numbers from Ref.~\cite{weiler} adjusted
  for all electron and muon final states, without ISR/FSR or
  hadronization or a complete fast detector simulation. All rates are
  given at leading order, to allow for a comparison with the original
  numbers in the last column.}
\label{tab:semilep}
\end{table}

The key observable shown in Table~\ref{tab:semilep} is the
signal-to-background ratio $S/B$, which determines how well we need to
know the theory and the systematics of the QCD backgrounds to extract
the signal. Note that none of the analyses shown offers a clear
side-bin background normalization. While the optimized analysis has an
increased signal efficiency by almost a factor ten and a promising
Gaussian statistical significance of $S/\sqrt{B} = 7.5$ (for
$10~\ifb$), values around $S/B \sim 1/7$ are clearly insufficient to
convincingly extract the stop signal in the presence of systematic and
theory errors.

The background results in Table~\ref{tab:semilep} should still be
taken with a grain of salt. While our signal efficiencies are in good
agreement between {\sc Pythia} and {\sc Herwig} (Fortran and C++), the
background numbers are sensitive to the underlying event. We can check
this effect by turning on/off the multi-parton interactions in {\sc
  Herwig++}, which leads to a decrease of the background rejection by
an order of magnitude. However, this does not affect the conclusion of
this section, namely that semi-leptonic stop searches are very
unlikely to be visible at the LHC. This is a generic statement in the
sense that looking at the systematic uncertainties we need to overcome
a relative factor of $\mathcal{O}(200)$ between the stop signal and
the top background rates and to our knowledge there is no kinematic
cut which for generic mass spectra significantly improves this ratio
after including detector smearing and fakes~\cite{ayres}.

\section{Hadronic Fat-Jet Analysis}
\label{sec:fat}

Given that the semi-leptonic analysis shown in the last section is
unlikely to work at all, an alternative strategy would be to search
for stop pairs in purely hadronic top decays. Those would allow us to
fully reconstruct the final state and analyze the angular correlation
in detail:
\begin{equation}
 pp \to \st1 \st1^*
    \to ( t \nz1) \, (\bar{t} \nz1) 
    \to ( b j j \nz1) \, (\bar{b} j j \nz1) \; .
\end{equation}
Our hadronic stop analysis is based on two tagged hadronic top quarks,
using the algorithm described in the Appendix. Tagging $W$ bosons in
their decays to geometrically large jets~\cite{fatjet_w} has been
around in the LHC literature for quite a while, including its
applications in searches for supersymmetry~\cite{fatjet_susy}. Higgs
tags can be implemented in a similar manner, and as it turns out they
show the best
performance~\cite{fatjet_vh,fatjet_tth,fatjet_susy,fatjet_higgs} when
based on the purely geometric C/A jet
algorithm~\cite{ca_algo,fastjet}. Inspired by searches for very heavy
resonances decaying to top pairs~\cite{tt_resonance} several top
taggers have been developed, again in the same spirit, but based on
different jet algorithms as well as on jet
shapes~\cite{top_tag,david_e,fatjet_tth,fatjet_wash}. One disadvantage
of most of these top taggers is that they are not designed to work for
the kind of transverse momenta we can expect in Standard Model
processes. This means that unlike the $W$ and Higgs
taggers~\cite{fatjet_vh,giacinto}, top taggers might be very hard to
establish experimentally. Following the $t\bar{t}H$
analysis~\cite{fatjet_tth} we slightly refine our top tagger for
moderate top boosts and apply it to this new challenge: extract a
new-physics signal from purely hadronic final states and reconstruct
its kinematics.\medskip

For triggering we expect our signal events to pass the jets plus
missing energy trigger at the LHC. To extract it from the backgrounds
we can employ the recently developed fat-jet tools which aim at
tagging a boosted top jet without being killed for example by
combinatorics. We start by constructing jets using the
Cambridge/Aachen algorithm~\cite{ca_algo}, implemented in {\sc
  Fastjet}~\cite{fastjet}, with $R=1.5$ and requiring at least two
jets with
\begin{equation}
p_{T,j} > 200/200~\gev \qqquad \qqquad
\met > 150~\gev \; .
\label{eq:cuts1}
\end{equation}
Those two cuts are chosen to obtain the largest signal-to-background
ratio $S/B$.  To reduce the probability of fake missing energy due to
detector effects we require the two-dimensional missing energy vector
to be well separated from the jets, to avoid cases where missing
energy is generated by just mis-measuring one jet. This should leave
us with a suppression factor of 1\% for fake missing energy above
$\met > 150$~GeV in QCD jet events without any physical missing
energy~\cite{csc_notes}, which we apply in the following.  Next, we
veto isolated leptons with $p_{T,\ell} > 15$~GeV, $|\eta_\ell|<2.5$,
requiring $E_T^\text{had} < 0.1 E_T^\text{lep}$ within $R<0.3$ around
the lepton. 

At this level we apply the top tagger described later and in the
Appendix and require two tops to be identified and reconstructed.
Finally, after requiring one $b$ tag inside the first tagged top we
construct $m_{T2}$~\cite{mt2}. Assuming we do not know the LSP mass,
\ie setting it to zero in the $m_{T2}$ construction, we require
\begin{equation}
m_{T2} > 250~\gev \; .
\end{equation}
While in Table~\ref{tab:had} we will see that this cut has hardly
any impact on the signal significance $S/\sqrt{B}$, at least for
small stop masses, we apply it to increase the signal-to-background
ratio $S/B$ and hence become less sensitive to systematic and theory
errors.

Constructing the $m_{T2}$ distributions has two motivations, of which
the background rejection cut might even be the lesser.  From the two
panels of Figure~\ref{fig:mt2} we see that $m_{T2}$ with an assumed
massless LSP is better suited to distinguish the stop signal from the
top background. As expected, Figure~\ref{fig:mt2} also shows that for
larger stop masses this cut becomes increasingly effective. More
importantly, once we know the correct value of $\mne1$ we can
determine the stop mass from the endpoint of the $m_{T2}$
distribution. Determining the uncertainties of such a mass
measurement, however, is beyond the scope of our phenomenological
analysis.  Obviously, due to the wrong decay topology the endpoint of
the $t\bar{t}$ background has nothing to do with the physical top
mass, so we cannot use it to gauge the stop mass measurement.\medskip

\begin{figure}[t]
\includegraphics[height=0.30\textwidth]{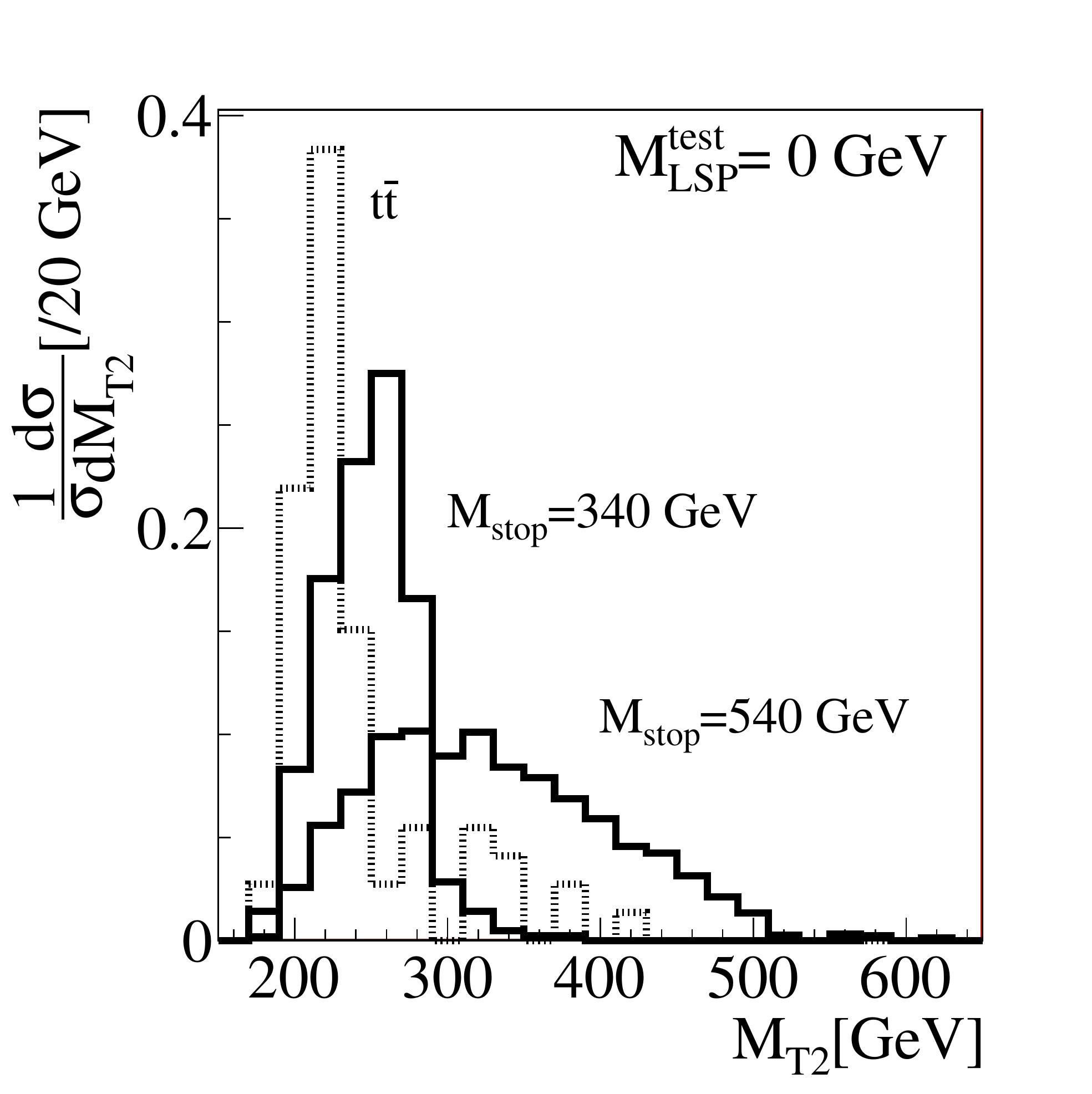}
\hspace*{0.2\textwidth}
\includegraphics[height=0.30\textwidth]{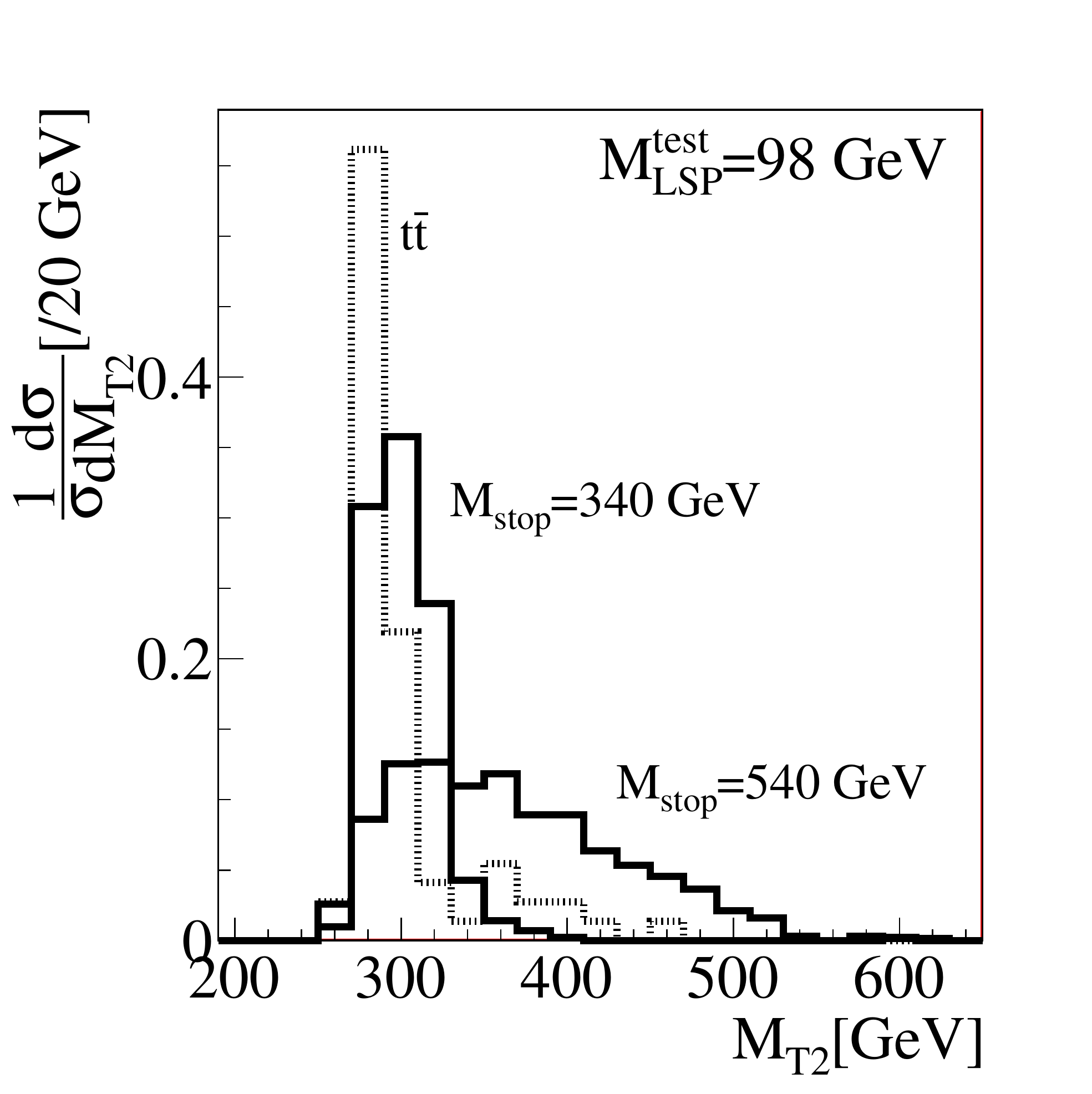}
\caption{Normalized $m_{T2}$ distributions for the stop signal ($\mst
  = 340$~GeV) and the $t\bar{t}$ background, after reconstructing two
  (real of fake) hadronic top quarks. The hypothetical LSP mass we set
  to $\mne1 = 0$~GeV (left) or to the correct value of $\mne1 =
  98$~GeV (right).}
\label{fig:mt2}
\end{figure}

For a double Standard Model top tag the mis-tagging probability when
applied to a pure QCD or $W$+jets sample after our process specific
cuts turns out to be (not much) below 0.1\%, comparable to the numbers
quoted in the Appendix, Table~\ref{tab:app_1_2}.  From the first
column of Table~\ref{tab:had} it is clear that such a reduction rate
is not sufficient.  Therefore, we follow the example of the Higgs
tagger~\cite{fatjet_vh,fatjet_tth} and apply an additional $b$ tag
inside the main constituents of the first tagged top. Limiting this
$b$ tag to the three main constituents of one specific tagged top
reduces the fake rate in particular from charm jets or gluons
splitting into $b\bar{b}$ pairs. Assuming a 60\% tagging efficiency
and a light-flavor rejection around 1/50 this will give the first top
tag a mistag rate well below 0.1\%. As it will turn out, this is
sufficient to render the QCD and $W$+jets backgrounds negligible
compared to the $t\bar{t}$ background. Charm jets in the QCD jets
sample we do not expect to be a problem. On the one hand, they have a
10\% mis-tagging probability for our $b$ tag, but on the other hand
the will appear much less frequently, based for example on the reduced
probability of gluon jets splitting into quarks --- a factor 1/4 from
counting quark flavors in $g \to q \bar{q}$ alone. Last but not least,
given the moderate boost of the top quarks we check that including a
$(0.1,0.1)$ granularity of the detector in a lego plot has no impact
on our analysis.

The large transverse momentum of the two candidate fat jets in
Eq.(\ref{eq:cuts1}) allows us not to worry about triggering on the one
hand and to generate events with a sizeable efficiency --- for the
actual analysis this cut has little effect, because inside the top
tagger we apply a lower cut on the transverse momentum of the
reconstructed top $p_{T,t}^\text{rec} > 200$~GeV. We explicitly check
this by lowering the acceptance cuts to $p_{T,j} > 100$~GeV and find
no effect on the final numbers of the analysis. \medskip

The different steps of our analysis are illustrated in
Table~\ref{tab:had}, for different stop masses and the leading
backgrounds. The event numbers are normalized to NLO cross sections
for the stop pair signal and the leading $t\bar{t}$ background. For
QCD and $W$+jets the normalization after cuts has many sources of
mostly experimental uncertainty that we can as well stick to the
leading order normalization.\medskip

\noindent
\underline{Stop pair signal} --- in contrast to for example Higgs
signals strongly interacting new particles will be produced with
sizeable rates at the LHC. For identical masses, the production rate
for stop pairs is actually the smallest of all QCD-initiated
supersymmetric processes, due to the large number of essentially
degenerate light-flavor squarks at the LHC, the fundamental color
charge of the stops, and the lack of a $t$-channel $q\bar{q}$
production process.  Typical NLO cross sections for stop pair
production range from 5.1~pb ($\mst = 340$~GeV) to 0.4~pb (540~GeV)
and 0.15~pb (640~GeV)~\cite{prospino}. After requiring missing energy,
two top tags and one $b$ tags we are left with several fb of rate. As
we can see in Figure~\ref{fig:mt2} the stiff $m_{T2}$ cut is not
particularly efficient, in particular for small stop masses, but it
does give us the necessary handle to suppress the $t\bar{t}$
background to a level of $S/B \sim \mathcal{O}(1)$.\smallskip

\noindent
\underline{Top pair background} --- as we know from the semi-leptonic
analysis and as we can see in Table~\ref{tab:had}, top pair production
is the most dangerous background to stop searches. Its total rate
shows a relative enhancement of several hundred over the signal and
two physical tops can be tagged including the $b$ jet we are
requiring. Purely hadronic top decays are reduced by our missing
energy cut in analogy to the pure QCD background, \ie by a factor
1/100. Semi-leptonic top decays are more dangerous, since after one top
tag the discussion in the Appendix shows that there is very likely a
second top tag based on recoiling QCD jets. After two top tags the
$t\bar{t}$ background is still larger than the signal. Therefore, we
apply a cut on $m_{T2}$, clearly distinguishing missing energy from
two LSPs to large missing energy from one neutrino in the semi-leptonic
top background.\smallskip

\noindent
\underline{QCD background} --- just because of its sheer size QCD jet
production tends to be an unsurmountable background at the LHC. After
requiring two hard jets we are still left with more than $10^7~\fb$ of
rate. As discussed in the Appendix we cannot suppress such a rate only
using the kinematic features of the top decay. The probabilistic
treatment of fake missing energy (1/100) and one $b$ mis-tag (1/50)
give us an additional suppression, where after two top tags we arrive
below the $t\bar{t}$ background. Note that we cannot assign a $m_{T2}$
survival probability to the QCD background, since we do not know the
distribution of the detector-fake missing energy vector. However,
because this fake missing energy will be uncorrelated with the other
momenta in the event, just like one additional missing particle, we
estimate the efficiency by the $t\bar{t}$ value around 22\%. If for
some reason QCD jet production should still pose an experimental
problem there is the option of requiring a $b$ tag also in the second
reconstructed top jet.\smallskip

\noindent
\underline{$W$+jets background} --- in contrast to QCD jets production
this process includes actual missing energy. Technically, we simulate
this background using {\sc Alpgen}~\cite{alpgen} with four hard jets
plus additional collinear jet radiation.  The $W$+jets rate only
exceeds the signal rate by less than a factor 100, so applying the
basic cuts and requiring two tagged top quarks reduces it to a level
we can deal with. The $b$ tag and the additional cut on $m_{T2}$
reduce the $W$+jets background to a level where it is hard to predict
without sufficient statistics. Irrespective of the details we can
conclude that $W$+jets do not pose a problem to the stop pair search.\smallskip

\noindent
\underline{$Z$+jets background} --- because of the significantly
smaller rate, the slightly lower invisible $Z$ branching ratio and the
sizeable probability to miss the lepton from the $W$ decay we can
safely assume that the $Z$+jets background will be as irrelevant as
the $W$+jets background after cuts. Numerically, even with too low
statistics for a detailed analysis we see that after cuts the $(Z \to
\nu\nu)$+jets background is always smaller than the $W$+jets
background by a factor $\mathcal{O}(1/3)$ and hence
irrelevant.\medskip

\begin{table}
\begin{tabular}{l|rrrrrr|rrrr||cr}
\hline 
& \multicolumn{6}{c|}{$\st1 \st1^*$}
& $t\bar{t}$ 
& QCD 
& $W$+jets 
& $Z$+jets 
& $S/B$ 
& $S/\sqrt{B}_{10~\ifb}$ \\
\hline 
$\mst[\gev]$ & 340 & 390 & 440 & 490 & 540 & 640 &&&&&
\multicolumn{2}{c}{340} \\
\hline
$p_{T,j}>200$~GeV, $\ell$ veto
& 728  & 447 & 292 & 187 & 124 & 46  & 87850 & $2.4 \cdot 10^7$ & $1.6 \cdot 10^5$ & n/a  & $3.0 \cdot 10^{-5}$ & \\
$\met > 150$~GeV 
& 283  & 234 & 184 & 133 & 93  & 35  &  2245 & $2.4 \cdot 10^5$ & 1710             & 2240 & $1.2 \cdot 10^{-3}$ & \\
first top tag
& 100  & 91  & 75  &  57 & 42  & 15  &   743 & 7590             & 90               & 114  & $1.2 \cdot 10^{-2}$ & \\
second top tag 
& 15   & 12.4& 11  & 8.4 & 6.3 & 2.3 &    32 &  129             & 5.7              & 1.4  & $8.3 \cdot 10^{-2}$ & \\
\hline
$b$ tag  
&8.7   & 7.4 & 6.3 & 5.0 & 3.8 & 1.4 &    19 & 2.6              &$\lesssim 0.2$  & $\lesssim 0.05$ & 0.40 & 5.9 \\
$m_{T2} > 250$~GeV
& 4.3  & 5.0 & 4.9 & 4.2 & 3.2 & 1.2 &   4.2 & $\lesssim 0.6$ &  $\lesssim 0.1$ &  $\lesssim 0.03$ & 0.88 & 6.1 \\
\hline
\end{tabular}
\caption{Signal (for different stop masses) and backgrounds for the
  hadronic fat-jet analysis. All numbers given in fb, the significance
  is computed for $10~\ifb$. The $\st1 \st1^*$ and $t\bar{t}$ rates
  are normalized to their higher-order
  values~\cite{top_rate,prospino}. $Z$+jets we simulate with the
  neutrino decay specified.}
\label{tab:had}
\end{table}

The right columns of Table~\ref{tab:had} clearly show that extracting
hadronic stop pairs from the different Standard Model backgrounds will
not be a problem at all. The statistical significance is above the
discovery limit already with an integrated luminosity of
$10~\ifb$. The event numbers are not huge, but a more careful
statistical treatment for example of our crude $m_{T2}$ cut will
change this easily. In contrast to semi-leptonic stop decays
systematics will not pose any problem either, possible complications
from jet combinatorics should be automatically resolved by the top
tagger~\cite{fatjet_tth}.

\medskip

One curious feature we see once we increase the stop mass: for a
constant LSP mass the increase in the cut efficiencies actually
over-compensates the decrease in the stop production rate. This is
most obvious for the $m_{T2}$ cut shown in Figure~\ref{fig:mt2}, but
it also holds for example for the top tagging efficiency which
benefits from the increased stop-neutralino mass
difference. Table~\ref{tab:had} therefore does not give a good
estimate of the stop mass reach.  To answer this question we would
need to adjust the $\met$ and $m_{T2}$ cuts which as it stands are
optimized for 340~GeV stops.\medskip

Moreover, it is clear that from the endpoints of the $m_{T2}$
distributions we should be able to measure the stop mass (or better
the stop--neutralino mass difference) in this process. While making
this quantitative statement does not require any further work,
actually estimating the experimental error on stop mass measurements
using fat jets goes far beyond what we can do in this paper. We
therefore refrain from quoting any number for the stop mass
measurements and leave it at this statement and the encouragement for
a detailed experimental analysis including full detector
simulation. For supersymmetric parameter analyses such a measurement
would of course be hugely beneficial~\cite{lhc_ilc,sfitter}.

\section{Outlook}
\label{sec:outlook}

We have shown that while semi-leptonically decaying stops are unlikely
to be observed at the LHC, a fat-jet analysis should be able to
discover purely hadronically decaying stops with typical integrated
luminosities of $10~\ifb$ at 14~TeV. This is true for stop masses
above 340~GeV (for $m_\text{LSP} = 98$~GeV) and extends to stop masses
well above this range. The stop mass reach based on hadronic decays
can be extended more by scaling the different cuts with the
stop-neutralino mass difference. Moreover, our limiting factor is
somewhat inefficient cuts to improve $S/B$, so we expect this result
to improve significantly once modern statistical methods are
applied.\medskip

The dominant background after cuts and reconstruction is exclusively
$t\bar{t}$ production, which we can reduce to the $S/B \sim 1$
level. QCD jet production is suppressed to a small fraction of the
$t\bar{t}$ background, and $V$+jets backgrounds are negligible.  This
promising result relies on two tagged and reconstructed top quarks,
which in turn allow us to use $m_{T2}$ constructed from the top
momenta and the missing energy vector. Combinatorics are automatically
resolved by the top tagging algorithm.\medskip

The fact that we can reconstruct the top momenta should allow the LHC
to analyze in detail the nature of a top partner decaying to a top
quark and a dark matter agent. Moreover, because of the large
signal-to-background ratio $S/B = \mathcal{O}(1)$ we will be able to use the
endpoints of the $m_{T2}$ distribution to measure the stop mass once
we know the LSP mass. Determining the experimental uncertainties for
this mass measurement we have to leave to an experimental study
including a full detector simulation.\medskip

As shown in detail in the Appendix our HEPTopTagger algorithm is not
only well suited to detect stop pairs at the LHC. It can be tested in
Standard Model top pair production and it can be applied to a large
variety of problems where standard methods fail, for example due to
jet combinatorics. In one such application, high multiplicities of
final states from longer decay chains will be automatically
resolved. In the current form the top tagger relies on a
Cambridge/Aachen algorithm with a mass drop criterion and a set of
invariant mass constraints. Once we require a fat jet with $p_T >
200$~GeV our top tagging efficiency can reach the 40\% to 50\% range
for reasonably boosted tops with mis-tagging probabilities around a
few per-cent.\bigskip \bigskip


\acknowledgments{}

We are very grateful to Gavin Salam for introducing us to this
exciting topic and for teaching us how to do actual analyses. To Are
Raklev we are grateful for carefully reading the manuscript. Moreover,
we would like to thank Gregor Kasieczka, Sebastian Sch\"atzel, and
Giacinto Picquadio for their useful input on experimental issues. MS and DZ
gratefully acknowledge the hospitality of the Institut f\"ur
Theoretische Physik in Heidelberg. This work was supported in part by
the US Department of Energy under contract number DE-FG02-96ER40969.

\appendix

\section{HEPTopTagger: Boosted Tops in the Standard Model}
\label{sec:appendix}

\begin{figure}[b]
\includegraphics[height=0.30\textwidth]{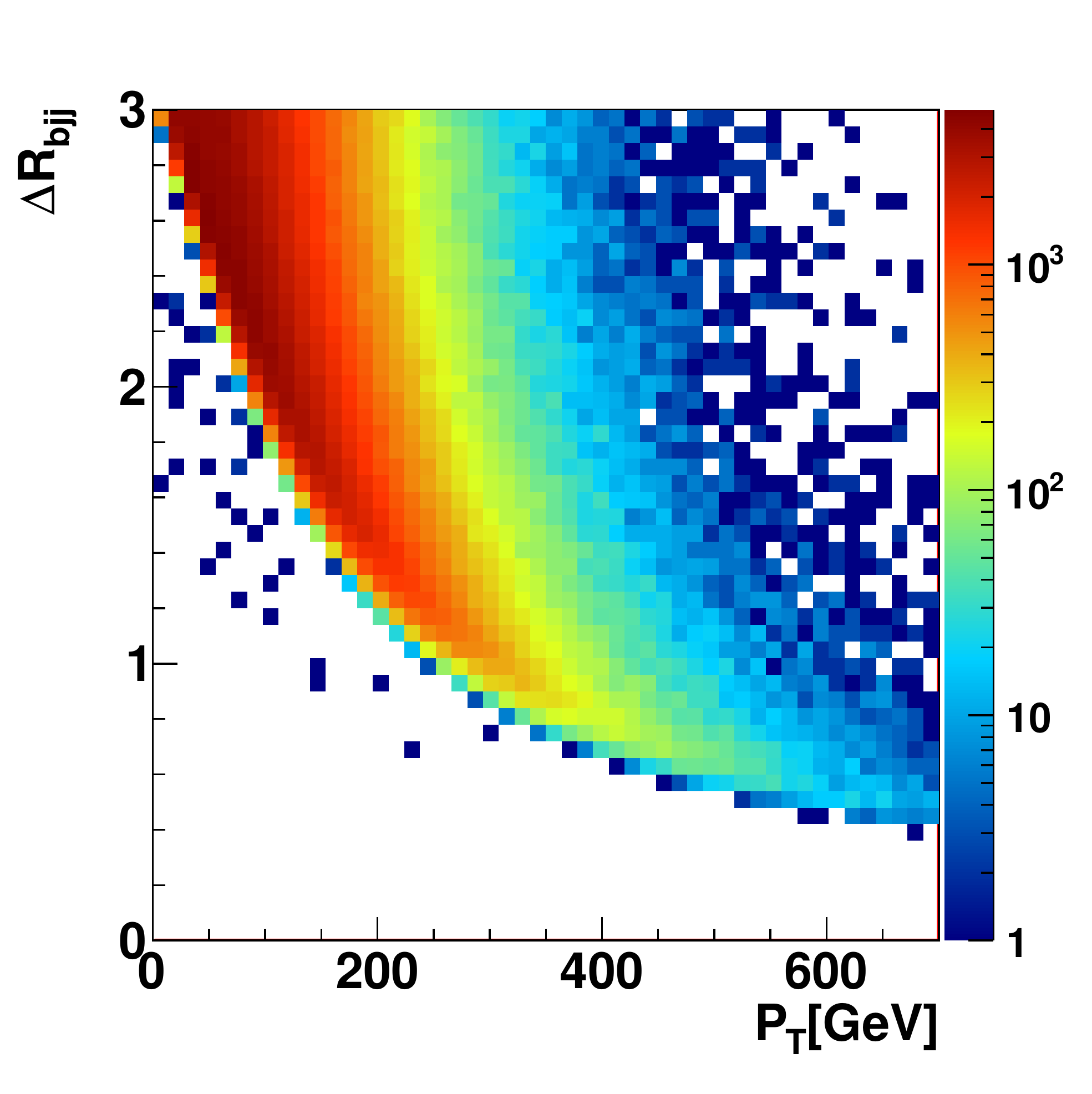}
\hspace*{0.2\textwidth}
\includegraphics[height=0.30\textwidth]{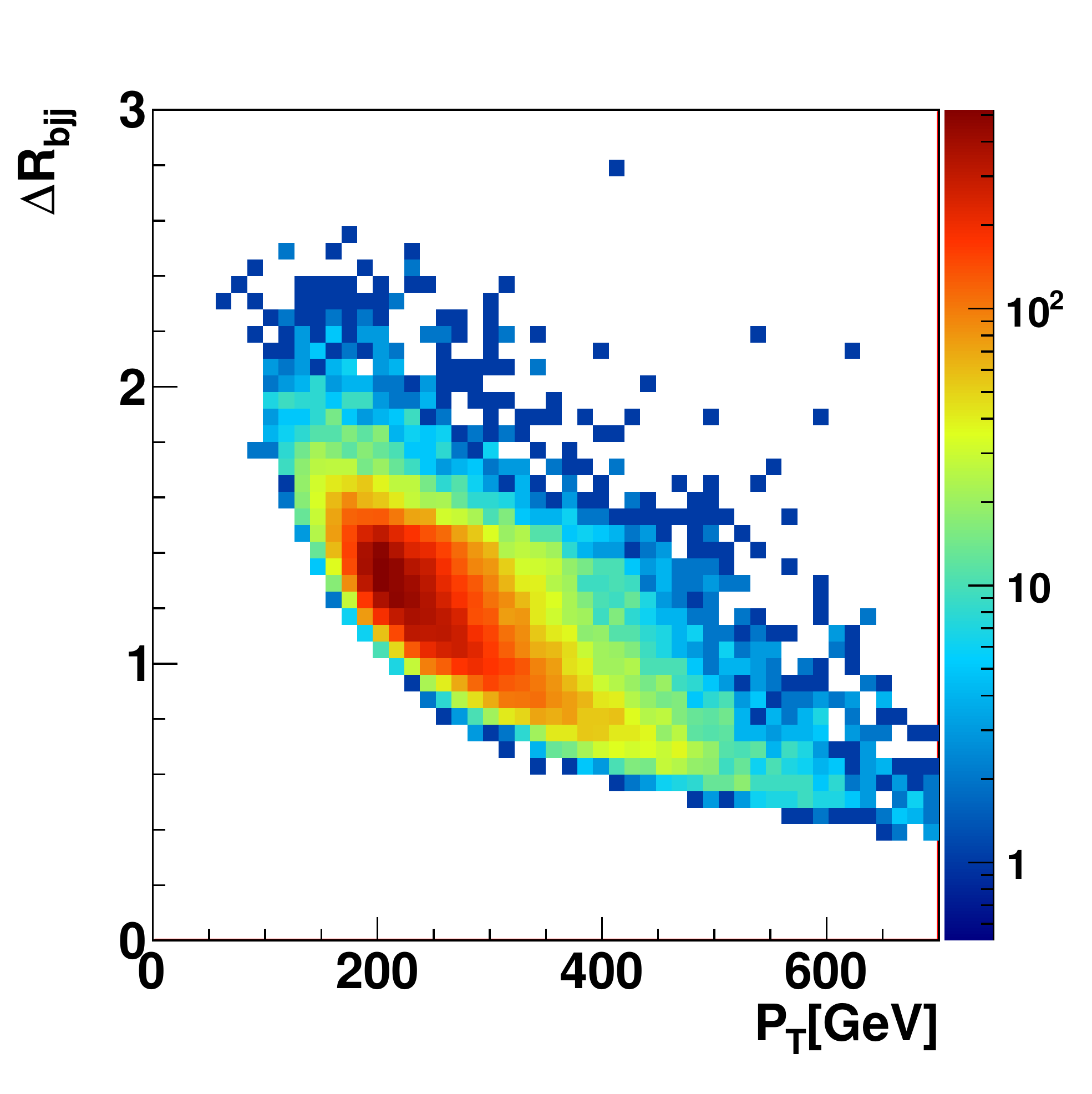}
\caption{Left: partonic $\Delta R_{bjj}$ vs $p_T$ distribution for a
  Standard Model $t\bar{t}$ sample. Right: the same correlation, but
  only for tagged top quarks and based on the reconstructed kinematic
  properties.}
\label{fig:app_drpt}
\end{figure}

Top taggers are algorithms identifying top quarks inside geometrically
large and massive jets. They rely on the way a jet algorithm combines
calorimeter towers into an actual jet. An obvious limitation is the
geometrical size of the jet which for a successful tag has to include
all three main decay products of the top quark. At the parton level we
can compute the size of the top quark from the three $R$ distances of
its main decay products: following the Cambridge/Aachen
algorithm~\cite{ca_algo,fastjet} we first identify the combination
$(i,j)$ with the smallest $\Delta R_{ij}$. The length of the second
axis in the top reconstruction we obtain from combining $i$ and $j$
and computing the $R$ distance of this vector to the third
constituent. The maximum of the two $R$ distances gives the
approximate partonic initial size $\Delta R_{bjj}$ of a C/A jet
covering the main top decay products. In Figure~\ref{fig:app_drpt} we
first correlate this partonic top size with the transverse momentum of
the top quark for a complete $t\bar{t}$ sample in the Standard
Model. As expected, if for technical reasons we want to limit the size
of the C/A fat jet to values below 1.5 we cannot expect to see top
quarks with a partonic transverse momentum of $p_T \lesssim
150$~GeV. In the right panel we show the same correlation, but after
tagging the top quark as described below and based on the
reconstructed kinematics. The lower boundaries indeed trace each
other, and the main body of tagged Standard Model top quarks resides
in the $p_{T,t}^\text{rec} = 200 \cdots 250$~GeV range, correlated
with $\Delta R_{bjj}^\text{rec} = 1 \cdots 1.5$. This result
illustrates that for a Standard Model top tagger it is indeed crucial
to start from a large initial jet size.\medskip

Therefore, our tagger for Standard Model tops is based on the
Cambridge/Aachen~\cite{ca_algo,fastjet} jet algorithm with $R=1.5$,
combined with a mass-drop
criterion~\cite{fatjet_vh,david_e,fatjet_tth}. Because the generic
$p_T$ range for the tops does not exceed 500~GeV the granularity of
the detector does not play a role, and we can optionally apply a $b$
tag to improve the QCD rejection rate. Since such a subjet $b$
tag~\cite{giacinto} will only enter as a probabilistic factor ($60\%,
10\%, 2\%$) for ($b,c,q/g$) jets we do not include it in the following
discussion. Note that whenever we require a $b$ tag in our actual
analysis, the numbers do not yet include the ($70\%, 1\%$)
improvements found for a $b$ tag inside a boosted
Higgs~\cite{giacinto}.\medskip

\begin{figure}[t]
\includegraphics[height=0.30\textwidth]{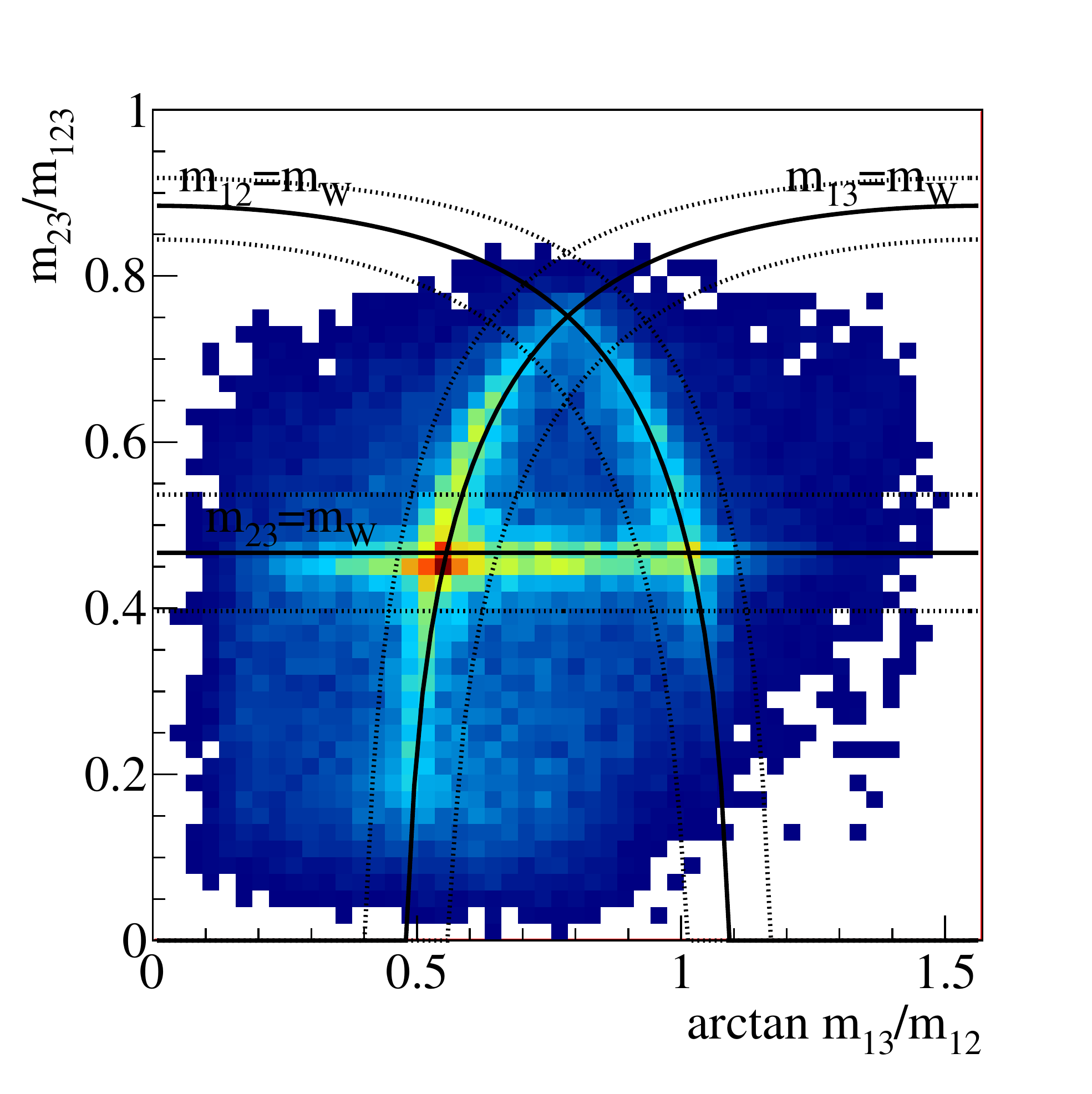}
\includegraphics[height=0.30\textwidth]{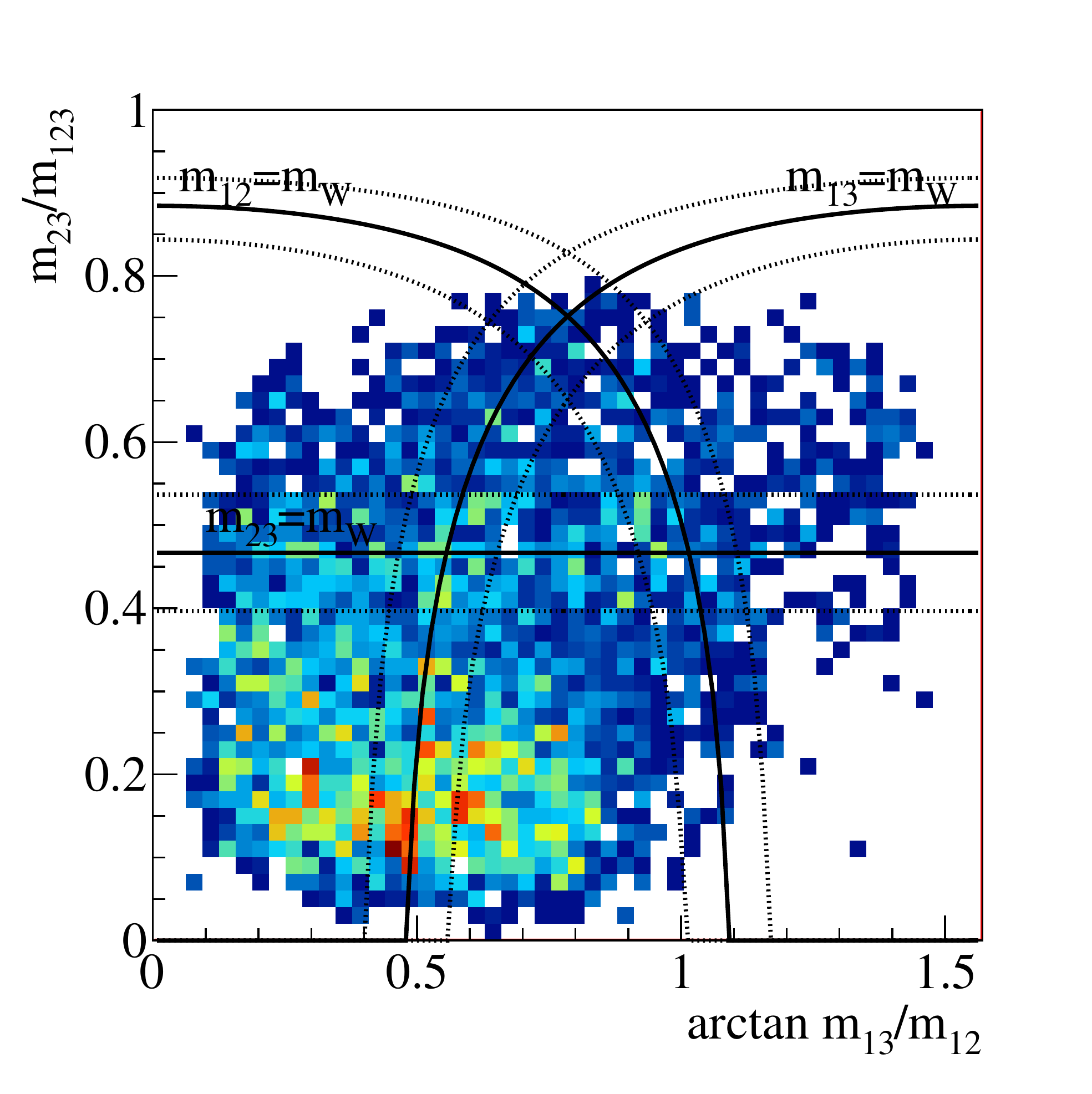}
\includegraphics[height=0.30\textwidth]{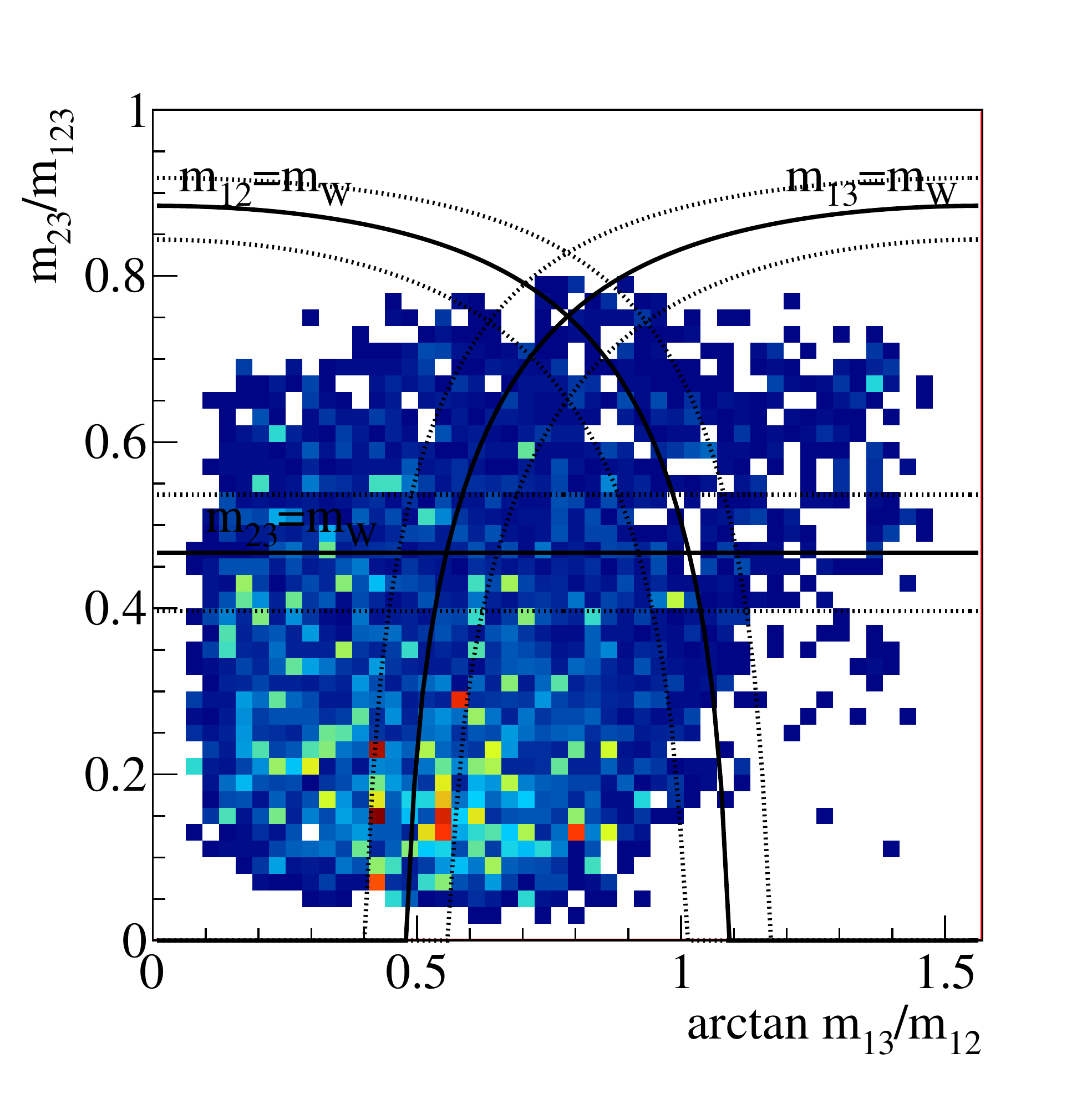}
\caption{Distribution of all events in the $\arctan m_{13}/m_{12}$ vs
  $m_{23}/m_{123}$ plane. We show $t\bar{t}$ (left). $W$+jets (center)
  and pure QCD jets (right) samples. More densely populated regions of
  the phase space appear in red.}
\label{fig:app_3d}
\end{figure}

The algorithm proceeds in the following steps:
\begin{enumerate}
\item define a fat jet using the C/A algorithm with $R=1.5$
\item for each fat jet, find all hard subjets using a mass drop
  criterion: when undoing the last clustering of the jet $j$, into two
  subjets $j_1,j_2$ with $m_{j_1} > m_{j_2}$, we require $m_{j_1} <
  0.8~m_j$ to keep $j_1$ and $j_2$. Otherwise, we keep only
  $j_1$. Each subjet $j_i$ we either further decompose (if $m_{j_i} >
  30~\gev$) or add to the list of relevant substructures.
\item iterate through all pairings of three hard subjets: first,
  filter them with resolution $R_\text{filter}=\min(0.3,\Delta
  R_{jk}/2)$.  Next, use the five hardest filtered constituents and
  calculate their jet mass (for less than five filtered constituents
  use all of them).  Finally, select the set of three-subjet pairings
  with a jet mass closest to $m_t$.
\item construct exactly three subjets $j_1, j_2, j_3$ from the five
  filtered constituents, ordered by $p_T$. If the masses $(m_{12},
  m_{13},m_{23})$ satisfy one of the following three criteria, accept
  them as a top candidate:
\begin{alignat}{5}
&0.2 <\arctan \frac{m_{13}}{m_{12}} < 1.3
\qquad \text{and} \quad
R_{\min}< \frac{m_{23}}{m_{123}} < R_{\max}
\notag \\
&R_{\min}^2 \left(1+\left(\frac{m_{13}}{m_{12}}\right)^2 \right) 
< 1-\left(\frac{m_{23}}{m_{123}} \right)^2
< R_{\max}^2 \left(1+\left(\frac{m_{13}}{m_{12}}\right)^2 \right)  	
\quad \text{and} \quad 
\frac{m_{23}}{m_{123}} > 0.35
\notag \\
&R_{\min}^2\left(1+\left(\frac{m_{12}}{m_{13}}\right)^2 \right) 
< 1-\left(\frac{m_{23}}{m_{123}} \right)^2
< R_{\max}^2\left(1+\left(\frac{m_{12}}{m_{13}}\right)^2 \right)  	
\quad \text{and} \quad 
\frac{m_{23}}{m_{123}}> 0.35
\label{eq:app_select}
\end{alignat} 
  with $R_{\min}=85\% \times m_W/m_t$ and $R_{\max}=115\% \times
  m_W/m_t$. The numerical soft cutoff at 0.35 is independent of the
  masses involved and only removes QCD events. The distributions for
  top and QCD events we show in Fig.~\ref{fig:app_3d}. 
\item finally, require the combined $p_T$ of the three subjets to
  exceed 200~GeV.
\end{enumerate}
\medskip

In step~3 of the algorithm there exist many possible criteria to
choose three jets from hard subjets inside a fat jet. For example, we
can include angular information (the $W$ helicity angle) in the
selection criterion and select the smallest $\Delta m_t + A_W \Delta
m_W + A_h \Delta \cos_h$.  In that case, the tagging efficiency
increases, but simultaneously the fake rate also increases, so to
reach the best signal significance we simply select the combination
with the best $m_t$. This allows us to apply efficient orthogonal
criteria based on the reconstructed $m_W$ and on the radiation pattern
later.\medskip

In step~4, the choice of mass variables shown in
Figure~\ref{fig:app_3d} is of course not unique. In general, we know
that in addition to the two mass constraints ($m_{123} =
m_t^\text{rec}$ as well as $m_{jk} = m_W^\text{rec}$ for one $(j,k)$)
we can exploit one more mass or angular relation of the three main
decay products. Our three subjets $j_k$ ignoring smearing and assuming
$p_i^2 \sim 0$ give
\begin{equation}
m_t^2 \equiv m_{123}^2 = (p_1 + p_2 + p_3)^2 = 
(p_1 + p_2)^2 + (p_1 + p_3)^2 + (p_2 + p_3)^2 =
m_{12}^2+m_{13}^2+m_{23}^2 \; ,
\end{equation}
which is the surface of a sphere with radius $m_t$ in $(m_{12},
m_{13}, m_{23})$. For fixed $m_{123}$ we can pick exactly two more
variables to fully describe the kinematics: we choose $m_{23}/m_{123}$
and $\arctan m_{13}/m_{12}$, which means that $m_{12}/m_{123}$ can be
derived as
\begin{equation}
1 = \left(\frac{m_{12}}{m_{123}}\right)^2
    \left( 1 + \left( \frac{m_{13}}{m_{12}}\right)^2\right)
    +\left(\frac{m_{23}}{m_{123}}\right)^2 \; .
\end{equation}
Assuming $m_{123}=m_t$ the condition $m_{12} = m_W \pm 15\%$ then
reads $m_{12}/m_{123} = R_\text{min} \cdots R_\text{max}$, which is
the form we use in Eq.(\ref{eq:app_select}). Note that our three mass
conditions can also be written in terms of two masses and the $W$
helicity angle~\cite{david_e,fatjet_tth}, but the construction of this
angle requires a boost into the $W$ rest frame with its experimental
challenges which we prefer to avoid. The switch from the helicity
angle scheme to the pure mass scheme only has a negligible effect on
the efficiencies computed without full detector simulation.\medskip
 
\begin{figure}[t]
\includegraphics[height=0.30\textwidth]{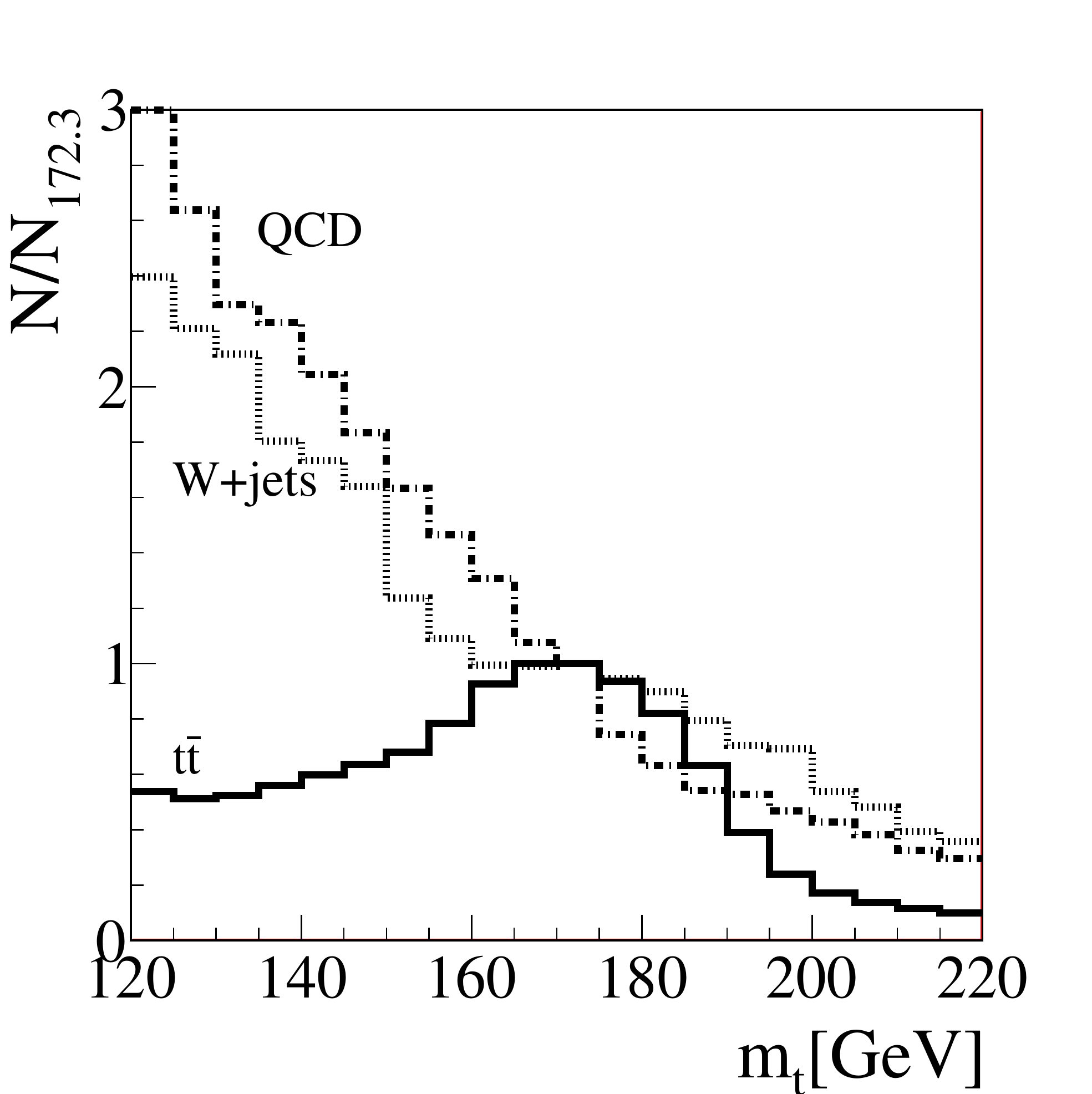}
\hspace*{0.2\textwidth}
\includegraphics[height=0.30\textwidth]{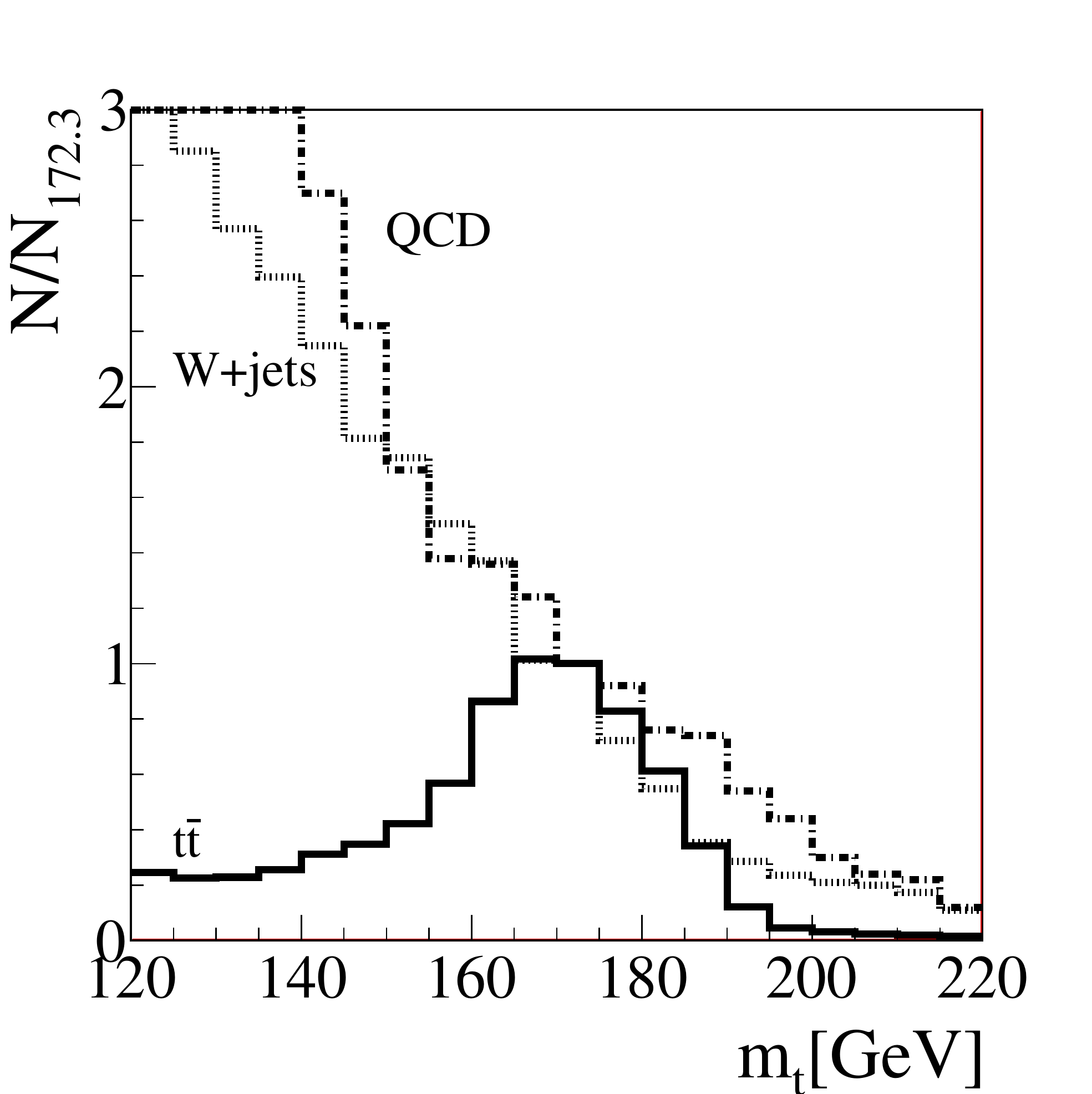}
\caption{Number of tagged tops in the $t\bar{t}$, QCD jets and
  $W$+jets channels for a varying assumed top mass in the tagging
  algorithm. The actual top mass in the sample is 172.3~GeV. Shown is
  the number of tagged first (left) and second (right) tops.}
\label{fig:app_sidebin}
\end{figure}

Finally, in contrast to the Higgs tagger~\cite{fatjet_vh, fatjet_tth}
the top tagger does know about the top mass when searching for the two
mass drops. This means that we will not be able to apply a side-bin
normalization. However, we can access side bins by changing the
assumed $m_t$ as used in the algorithm, Eq.(\ref{eq:app_select}), to
values different from the top mass in the event sample. The result of
such a misalignment we show in Figure~\ref{fig:app_sidebin}: for the
QCD and $W$+jets background the number of tagged tops follows the
typical $p_T$ dependence of the jet sample. The lower the top mass we
are looking for the more tagged tops we will find. In contrast, the
top sample shows a clear peak when the assumed top mass in the
algorithm coincides with the top mass in the sample. Towards larger
assumed top masses the distribution shows a one-sided width around
20~GeV. Towards smaller assumed top masses additional QCD jets can
have an increasing impact on wrongly tagged tops. Therefore, the tail
is considerably higher. While this kind of behavior makes it unlikely
that such side-bins will useful for an actual analysis they serve as a
very useful cross check for our fat-jet methods.\medskip

In Figure~\ref{fig:app_eff} we summarize the performance of the
tagging algorithm described above. In the left panel, we show the
parton-level $p_T$ of the hadronic top quarks in the $t \bar{t}$
sample, normalized to the top production rate. As we already know from
Figure~\ref{fig:app_drpt} this distribution drops rapidly and
essentially vanishes for $p_T > 500$~GeV. This is the reason why our
tagging algorithm focuses on a top $p_T$ range between 200 and
500~GeV. The curve for tagged tops follows the curve for produced tops
smoothly for $p_T > 250$~GeV. The same curve for tagged tops is
actually included in all three panels, different just because of the
normalization of the plots. The two curves for mis-tagged tops in the
$W$+jets and QCD sample are shown as a function of the reconstructed
$p_T$ of the top constituents in the last step of our
algorithm. Again, they are normalized to the production rate at the
LHC, so we immediately see that one top tag will not be sufficient to
reduce the pure QCD background to the level of top pair
production. The $W$+jets background, in contrast, should not pose a
problem to fat-jet analyses, which we confirm in our actual analysis
and show in Table~\ref{tab:had}.

\begin{figure}[t]
\includegraphics[height=0.30\textwidth]{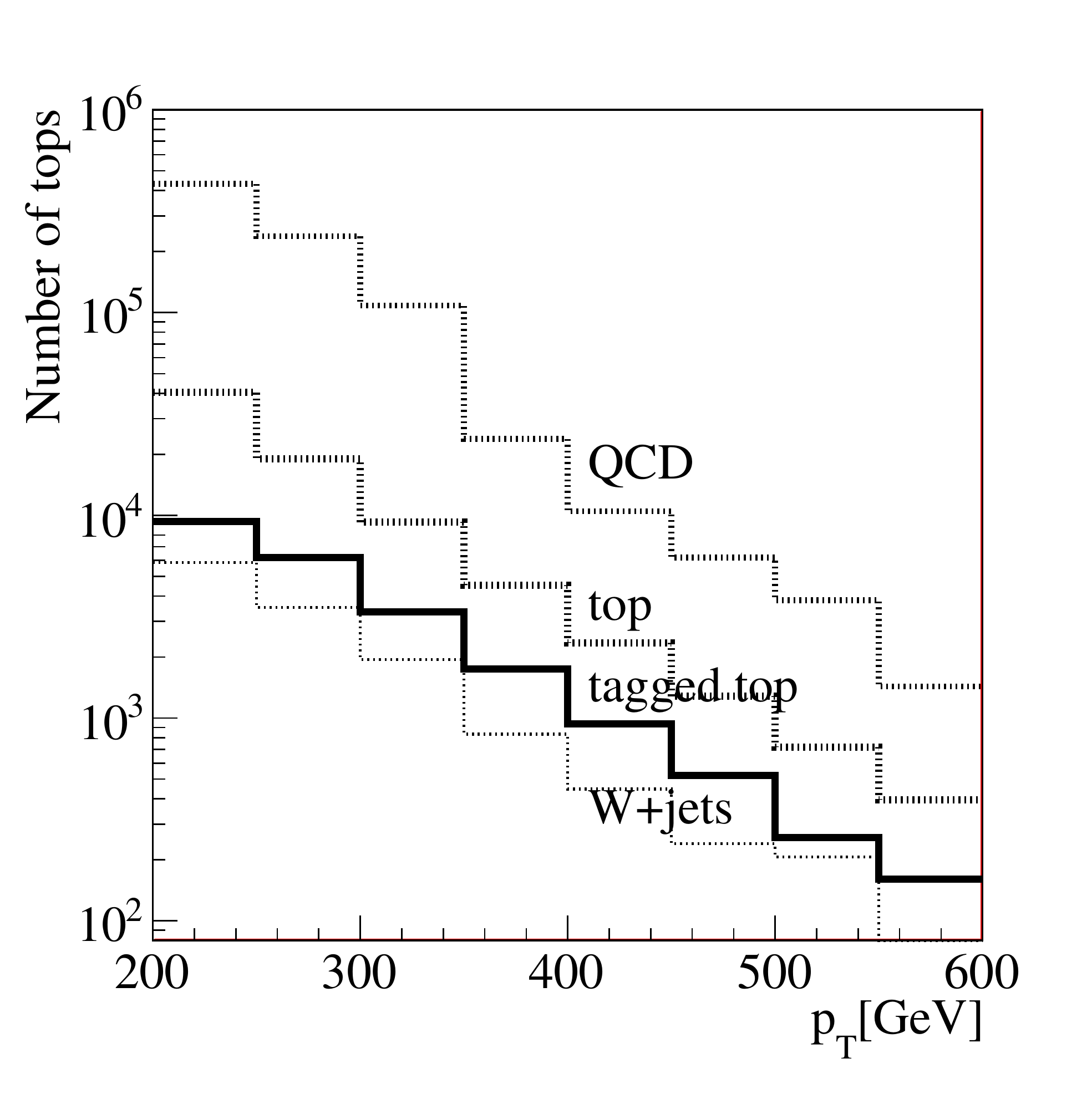}
\includegraphics[height=0.30\textwidth]{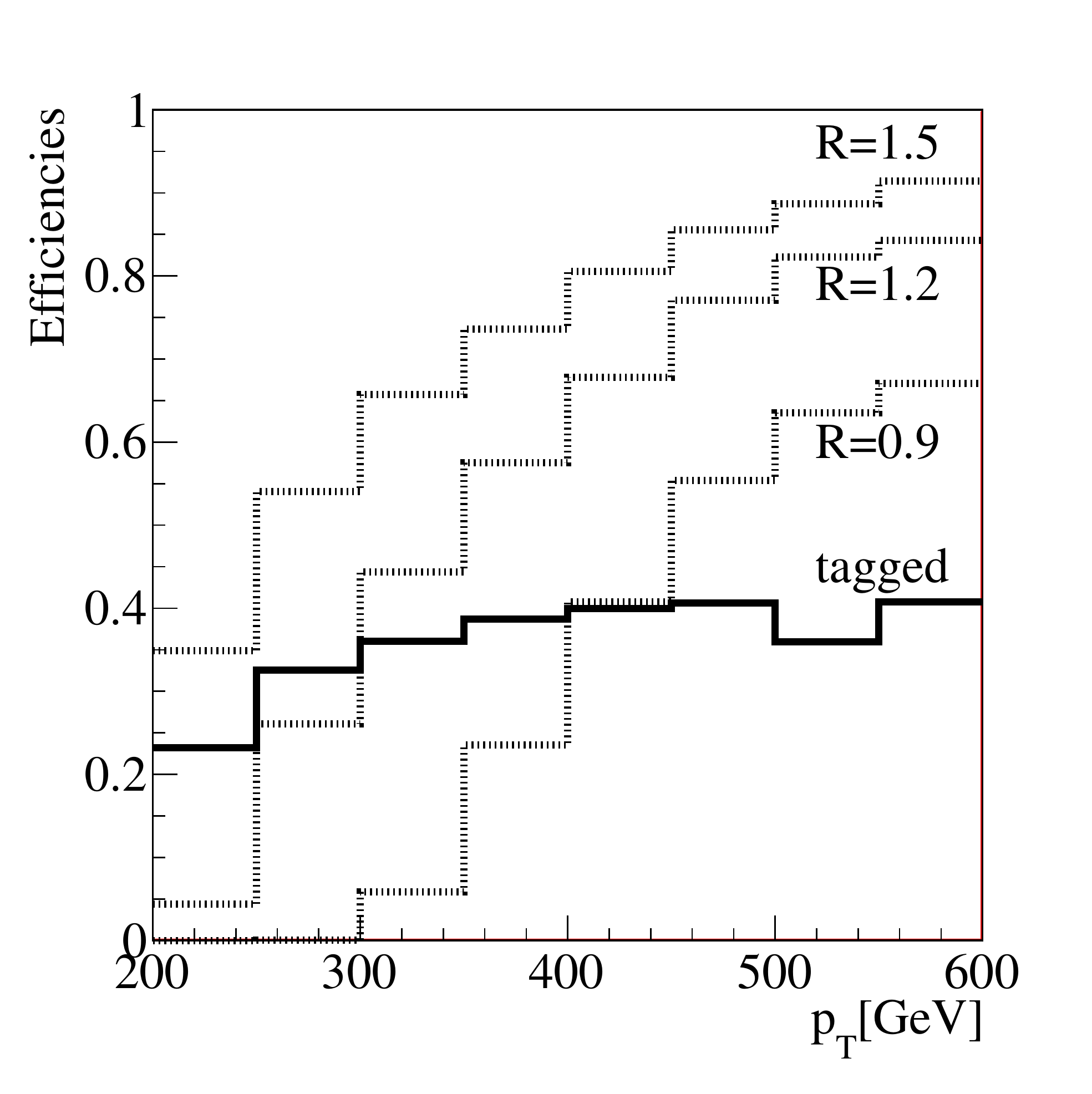}
\includegraphics[height=0.30\textwidth]{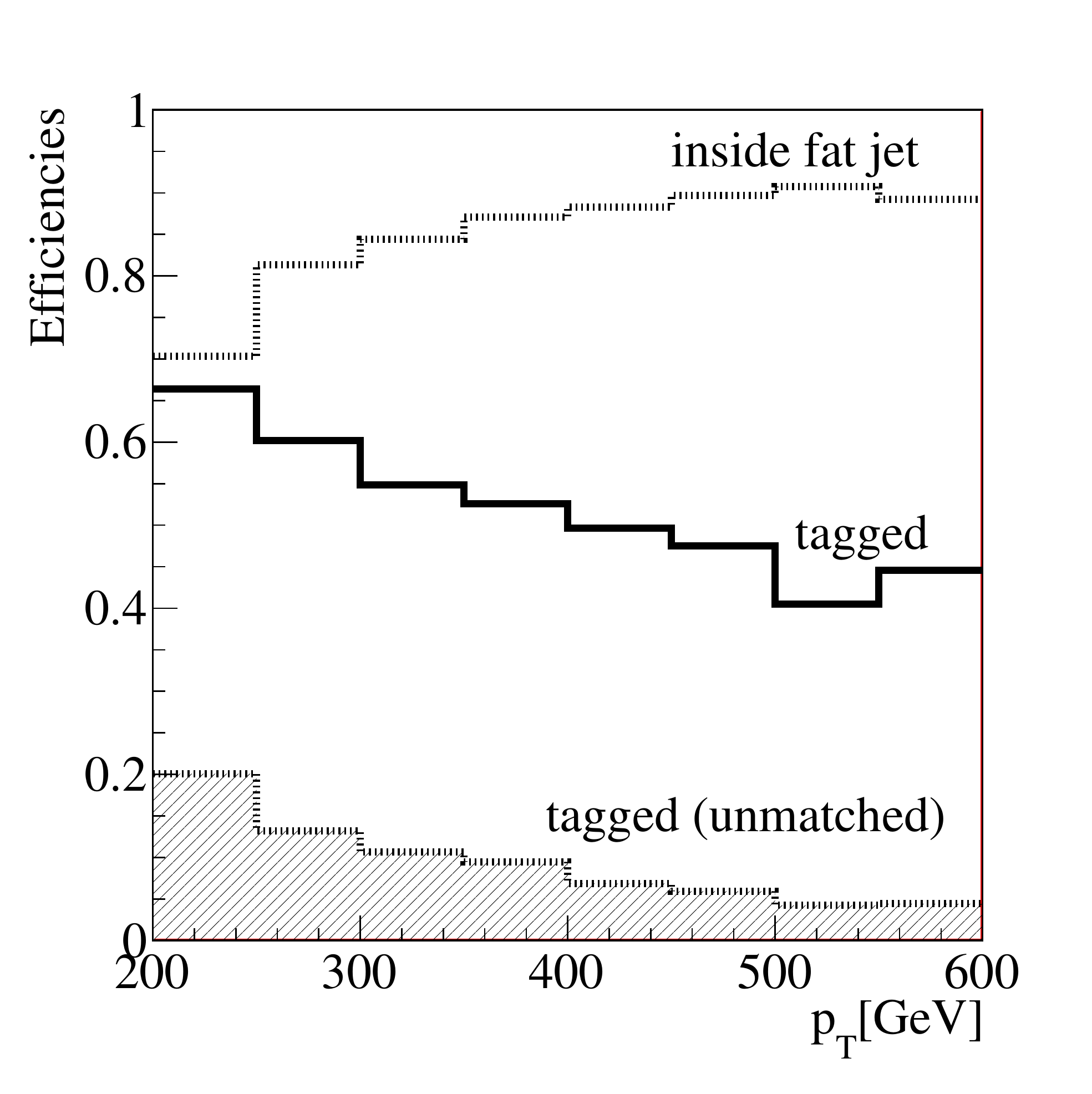}
\caption{Left: number of tops, tagged tops, mis-tagged tops from QCD
  jets and for $W$+jets for $1~\ifb$. Center: fraction of hadronic
  tops whose main parton-level decay products are within a C/A
  distance of $\Delta R_{bjj}=1.5, 1.2, 0.9$.  Right: tagging
  efficiencies, normalized to the top line of the central panel.}
\label{fig:app_eff}
\end{figure}

In the center panel of Figure~\ref{fig:app_eff} we show the fraction
of tops found inside C/A distances of $\Delta R_{bjj} < 0.9,1.2,1.5$,
normalized to the number of tops produced (\ie the top production line
in the left panel).  As indicated in Figure~\ref{fig:app_drpt}, in
particular in the promising Standard Model range $p_{T,t} < 300$~GeV
we lose the vast majority of events if we reduce the jet size from
$R=1.5$ to $R=1.2$. We also show the fraction of tagged tops based on
$R=1.5$, showing an efficiency of 20\% to 40\% relative to all tops
produced.

In the right panel of Figure~\ref{fig:app_eff} we show the top tagging
efficiency as a function of the reconstructed top $p_T$, normalized to
the number of tops within $\Delta R_{bjj}<1.5$ (the top line in the
center panel). The first line shows how many of the tops end up with
all main parton-level decay products inside the fat jet, as requested
in step~1 of our algorithm. There is a loss associated with this
actual construction of the fat jet, because even if all main top decay
products are close enough to end up inside a fat jet of size $R=1.5$,
they do not have to. For example, the geometric center of the fat jet
can be slightly shifted, so one of the top decay products drops out.
The second line shows the fraction of tagged tops. The shading
indicates the fraction of these tops where we cannot establish a
one-to-one connection between the three subjets constructed in step~4
and the parton-level top decay products. 

To establish such a connection we compute the $R$ distances between
the three subjets and all hard partons in the event. We then identify
the parton pairing which gives the smallest value of $\Delta R_{ijk}^2
= \Delta R(j_1,p_i)^2 + \Delta R(j_2,p_j)^2 + \Delta R(j_3,p_k)^2$ and
check if this pairing corresponds to a top decay at parton level. If
not, we assume that either a QCD jet might have entered the
reconstruction or that QCD radiation has bent one of the top decay
jets far away from its partonic origin. However, this rate is
considerably higher than the $W$+jets mis-tag rate, so these events
are not dominated by continuum QCD jet production. Instead, they
represent the generic problem of identifying partons with jets by some
kind of geometric measure.  In a way these tags are the tricky ones
for low transverse momenta, while the efficiency for identifiable tags
is a fairly constant $\mathcal{O}(40\%)$ over the entire $p_T$
range.\medskip

\begin{table}[b]
\begin{tabular}{l|rrrrr|rrrrr|l}
\hline 
& \multicolumn{3}{c}{$t\bar{t}$}  & QCD & $W$+jets 
& \multicolumn{3}{c}{$t\bar{t}$}  & QCD & $W$+jets  \cr
\hline
$p_{T,t}^\text{min} [\gev]$
& 0 & 200 & 300 &&
& 0 & 200 & 300 && \cr
\hline \hline
one fat jet 
& $92200$ & $36100$ & $8250$ & $4.10 \cdot 10^7$ & $3.19 \cdot 10^5$ 
& 100\% & 100\% &100\%  &100\% & 100\% & \cr 
two fat jets 
& $40700$ & $20300$ & $5810$ & $2.16 \cdot 10^7$ & $1.60 \cdot 10^5$
 & 44\%& 57\% & 70\% & 53\% & 50\% & relative to one fat jet \cr \hline
one top tag 
& $20900$ & $13400$ & $4160$ & $8.18 \cdot 10^5$ & $1.27 \cdot 10^4$ 
& 23\% & 37\% & 51\% & 2.0\% & 3.9\% & relative to one fat jet \cr
two top tags  
& $1880$ & $1630$ & $700$ & $11000$ & $233$            
 & 2.0\% & 4.5\% & 8.5\% & 0.027\% & 0.07\% & relative to one fat jet \cr
&&&&&                            
& 4.5\% &8.0\%  & 12\% & 0.05\% & 0.15\% & relative to two fat jets \cr
\hline
\end{tabular}
\caption{Number of events in $1~\ifb$ and their relative tagging
  efficiencies for the first and the second top tag in the different
  Standard Model event samples. One top tag and one fat jet means at
  least one tag or fat jet. Fat jets are defined with
  $p_T>200$~GeV. The top quarks are produced with $p_T >
  0,200,300$~GeV at parton level. All rates are quoted at leading
  order.}
\label{tab:app_1_2}
\end{table}

The bottom line in terms of tagging efficiencies and mis-tagging
probabilities we show in Table~\ref{tab:app_1_2}: provided we find
something like a fat jet a top can be tagged with an efficiency of
23\% to 51\%, dependent on the $p_T$ range of the top. This variation
shows that for low generated $p_{T,t}$ there will still be a fat jet
in the $t\bar{t}$ sample, but this fat jet will tend to not include
the top decay products, so we cannot tag a top to begin with.

For a second top tag we first need to see another fat jet in the
sample. For top pairs this will happen in 44\% of all events, up to
70\% for hard tops.  However, there are two tops in the event, and
there will likely be two fat jets. For $p_{T,t} > 200$~GeV the 37\%
tagging efficiency quoted in Table~\ref{tab:app_1_2} corresponds to a
26\% efficiency of tagging a top in a given fat jet. Based on this
number we can compute the probability of tagging two tops in two fat
jets, which gives slightly less than 4\%.  So in particular for
low-$p_T$ tops our efficiency for a second top tag is higher than for
the first.  For the signal discussed in the main body of the paper we
would need to fold this efficiency with the $p_T$ spectrum of tops
from stop decays. From the 200~GeV and 300~GeV columns in
Table~\ref{tab:app_1_2} we see that this will help considerably.

For $W$+jets and pure QCD jets with their generically softer QCD
structure it will not be as likely to actually find the first fat jet
in the sample. In addition, the mis-tagging rate for the first top tag
after seeing a fat jet ranges around 2\% to 4\%.  The efficiency for a
second fat jet in the background processes is almost as large as for
top pairs.  This reflects the fact that one hard fat jet has to recoil
against QCD activity which will give us a second fat jet. The
probability of mis-tagging two tops is then roughly the first mis-tag
probability squared, after factoring out the probabilities of finding
one or two fat jets.  This way, the over-all efficiencies for two top
tags significantly enhance the signal-to-background ratio, in
particular for pure QCD jets. As mentioned several times, this number
can be improved if we ask for a $b$ tag inside the top jet. As a last
comment, our top tagger is optimized for low $p_{T,t}$, so further
work and modifications should be able to increase its efficiency
towards higher boosts.\medskip

\begin{figure}[t]
\includegraphics[height=0.30\textwidth]{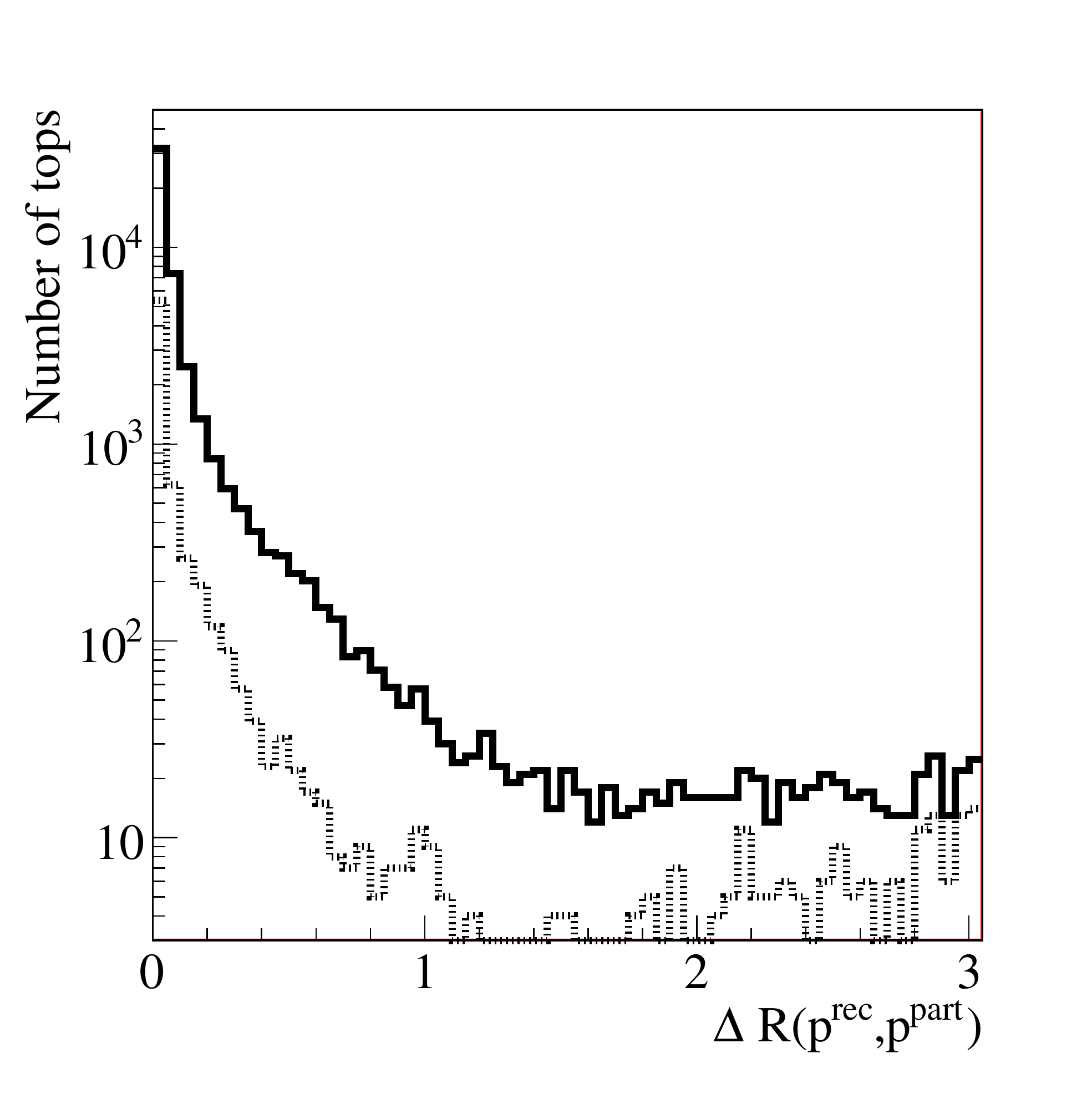}
\includegraphics[height=0.30\textwidth]{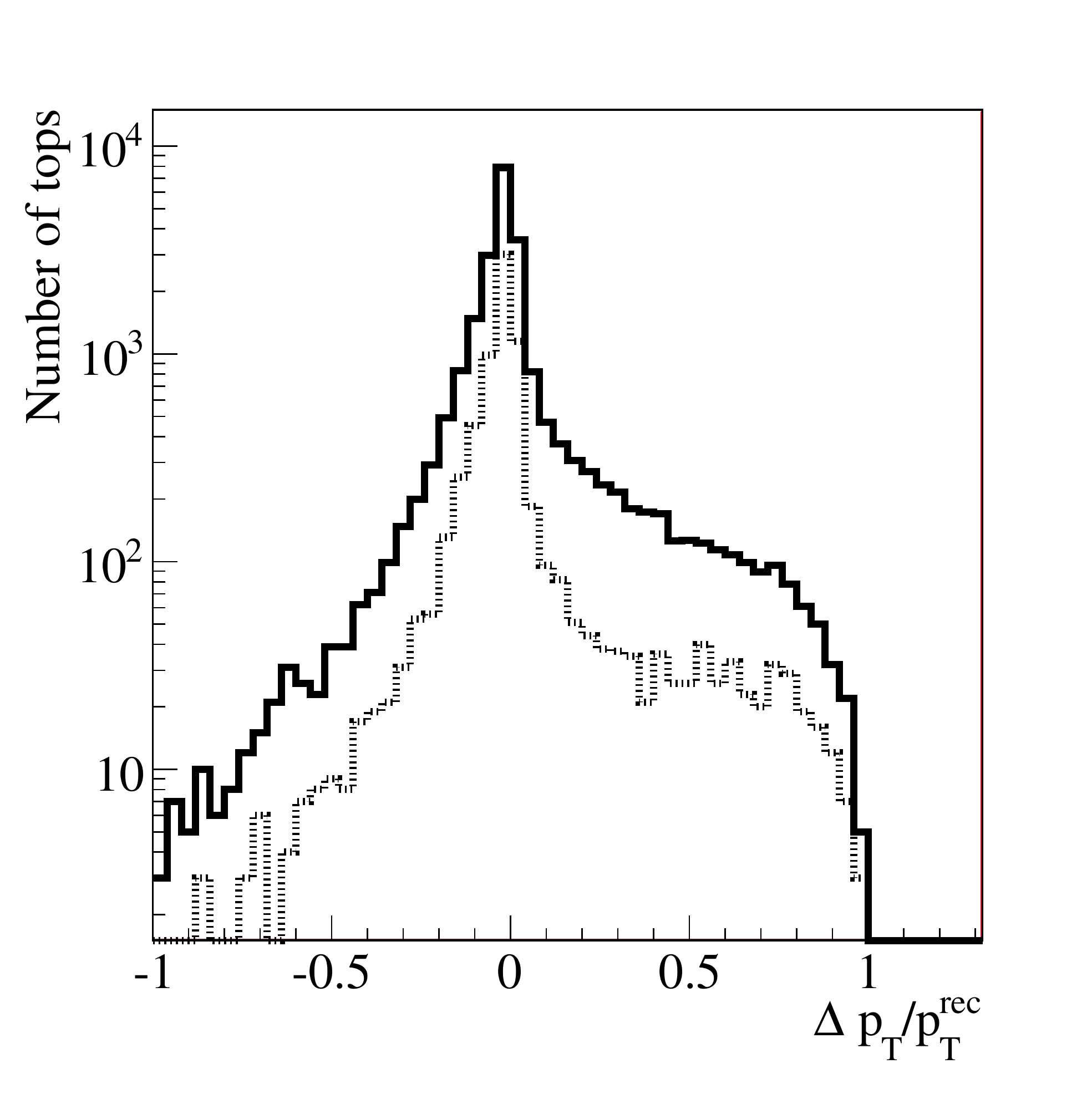}
\includegraphics[height=0.30\textwidth]{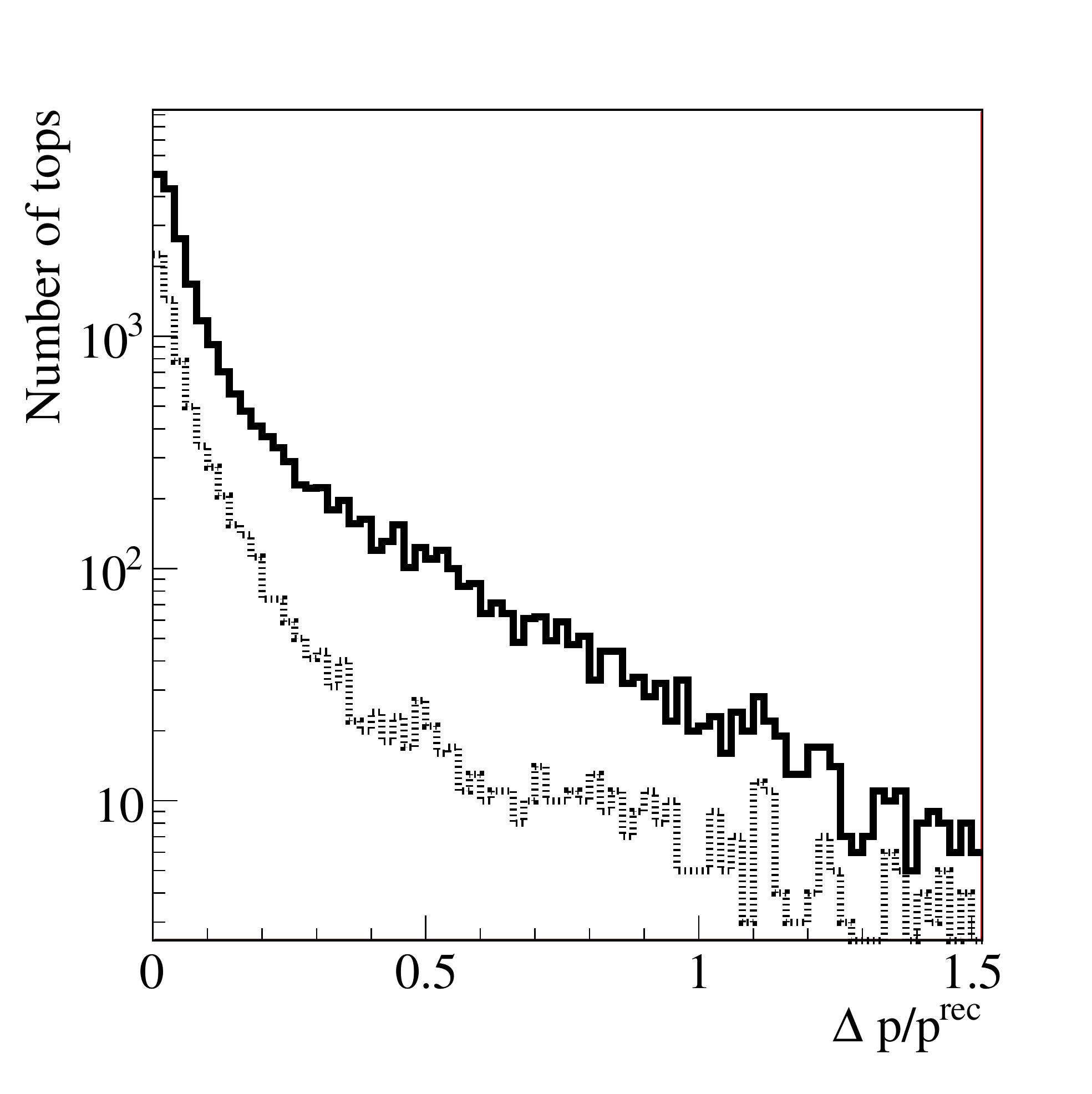}
\caption{From left to right: $\Delta R$ between the reconstructed and
  the parton-level top quark; $\Delta p_T/p_T^\text{rec}$ for the same
  sample; $\Delta p/p^\text{rec}$ where $p$ denotes the absolute value
  of the top 3-momentum.  The solid curves correspond to the default
  $p_{T,t}^\text{rec} > 200$~GeV of the tagged top, while the dashed
  curves constrains the tagged top to values above $p_{T,t}^\text{rec}
  > 300$~GeV.}
\label{fig:app_reconstruct}
\end{figure}

The last question beyond the simple top tag is how well the algorithm
described in this Appendix can reconstruct the top momentum.  In
principle, our top tagging algorithm can identify the three subjets as
either $W$-decay jets or the $b$ jet. Unfortunately, even amongst the
events which allow for a clear comparison of the partonic top decay
products and the resulting subjets a fair fraction returns $m_{bj}
\sim 80$~GeV on the parton level. This is because the invariant masses
of all jet combinations reside in the same range, out of which our $W$
mass window represents a sizeable fraction. To test for such effects
we can take all tagged tops in the hadronic $t\bar{t}$ sample and
compare the reconstructed top momenta to the parton-level input. In
Figure~\ref{fig:app_reconstruct} we first show the angular distance of
the reconstructed top quarks from the parton-level truth. 

While there is a strong peak for $\Delta R < 0.5$ and 95\% of the
events resides in the area, we also observe a long tail at the
$10^{-3}$ level, which is due to combinatorics or effective QCD
mis-tags in the top sample.  In the second panel we show the relative
error on the transverse momentum of the tagged top ($\Delta
p_T=p_T^\text{rec} - p_T^\text{part}$).  For around 85\% of the events
the mis-measurement compared to the parton-level truth stays below the
20\% level. In the third panel we show the same for the entire
3-momentum. Again, 68\% of the tops are reconstructed at the 10\%
level, while 80\% are reconstructed within $\Delta p/p \sim
20\%$. Obviously, all these numbers can be improved if we increase the
$p_T^\text{min}$ cut on the reconstructed top for example from 200~GeV
to 300~GeV.



\begin{thebibliography}{99}

\bibitem{review}
  D.~E.~Morrissey, T.~Plehn and T.~M.~P.~Tait,
  arXiv:0912.3259 [hep-ph].

\bibitem{meade}
  P.~Meade and M.~Reece,
  Phys.\ Rev.\  D {\bf 74}, 015010 (2006).

\bibitem{stop_decays}
 A.~Djouadi, W.~Hollik and C.~J\"unger,
  Phys.\ Rev.\  D {\bf 54}, 5629 (1996);
 S.~Kraml, H.~Eberl, A.~Bartl, W.~Majerotto and W.~Porod,
  Phys.\ Lett.\  B {\bf 386}, 175 (1996);
 W.~Beenakker, R.~H\"opker, T.~Plehn and P.~M.~Zerwas,
  Z.\ Phys.\  C {\bf 75}, 349 (1997).

\bibitem{stops_tevatron_c}
  T.~Aaltonen {\it et al.}  [CDF Collaboration], 
   CDF Note 9834, 2009;
  V.~M.~Abazov {\it et al.}  [D0 Collaboration],
  Phys.\ Lett.\  B {\bf 665}, 1 (2008).

\bibitem{stops_tevatron_b}
  T.~Aaltonen {\it et al.}  [CDF Collaboration],
  arXiv:0912.1308 [hep-ex];
  V.~M.~Abazov {\it et al.}  [D0 Collaboration],
  Phys.\ Lett.\  B {\bf 675}, 289 (2009).

\bibitem{cms_tdr}
 G.~L.~Bayatian {\it et al.}  [CMS Collaboration],
  J.\ Phys.\ G {\bf 34}, 995 (2007).

\bibitem{stops_measure}
  P.~Langacker, G.~Paz, L.~T.~Wang and I.~Yavin,
  JHEP {\bf 0707}, 055 (2007);
  J.~Ellis, F.~Moortgat, G.~Moortgat-Pick, J.~M.~Smillie and J.~Tattersall,
  Eur.\ Phys.\ J.\  C {\bf 60}, 633 (2009);
  K.~Rolbiecki, J.~Tattersall and G.~Moortgat-Pick,
  arXiv:0909.3196 [hep-ph];
  M.~Blanke, D.~Curtin and M.~Perelstein,
  arXiv:1004.5350 [hep-ph].

\bibitem{weiler}
  M.~Perelstein and A.~Weiler,
  JHEP {\bf 0903}, 141 (2009).

\bibitem{fatjet_vh}
  J.~M.~Butterworth, A.~R.~Davison, M.~Rubin and G.~P.~Salam,
  Phys.\ Rev.\ Lett.\  {\bf 100}, 242001 (2008);
  D.~E.~Soper and M.~Spannowsky,
  arXiv:1005.0417 [hep-ph].

\bibitem{david_e}
 D.~E.~Kaplan, K.~Rehermann, M.~D.~Schwartz and B.~Tweedie,
  Phys.\ Rev.\ Lett.\  {\bf 101}, 142001 (2008).

\bibitem{fatjet_tth}
  T.~Plehn, G.~P.~Salam and M.~Spannowsky,
  Phys.\ Rev.\ Lett.\  {\bf 104}, 111801 (2010).

\bibitem{prospino}
  W.~Beenakker, M.~Kr\"amer, T.~Plehn, M.~Spira and P.~M.~Zerwas,
  Nucl.\ Phys.\  B {\bf 515}, 3 (1998).

\bibitem{top_rate}
  P.~Nason, S.~Dawson and R.~K.~Ellis,
  Nucl.\ Phys.\  B {\bf 303}, 607 (1988);
  W.~Beenakker, H.~Kuijf, W.~L.~van Neerven and J.~Smith,
  Phys.\ Rev.\  D {\bf 40}, 54 (1989);
  S.~Moch and P.~Uwer,
  Phys.\ Rev.\  D {\bf 78}, 034003 (2008).

\bibitem{top_rec_semilep}
 see \eg 
 V.~Barger, T.~Han and D.~G.~E.~Walker,
  Phys.\ Rev.\ Lett.\  {\bf 100}, 031801 (2008);
 U.~Baur and L.~H.~Orr,
  Phys.\ Rev.\  D {\bf 76}, 094012 (2007);
 T.~Han, R.~Mahbubani, D.~G.~E.~Walker and L.~T.~E.~Wang,
  JHEP {\bf 0905}, 117 (2009).

\bibitem{herwig}
 G.~Corcella {\it et al.},
  arXiv:hep-ph/0210213;
 M.~Bahr \etal,
  arXiv:0812.0529 [hep-ph].

\bibitem{pythia}
  T.~Sjostrand, S.~Mrenna and P.~Z.~Skands,
  JHEP {\bf 0605}, 026 (2006);
  T.~Sjostrand, S.~Mrenna and P.~Z.~Skands,
  Comput.\ Phys.\ Commun.\  {\bf 178}, 852 (2008).

\bibitem{alpgen}
 M.~L.~Mangano, M.~Moretti, F.~Piccinini, R.~Pittau and A.~D.~Polosa,
  JHEP {\bf 0307}, 001 (2003).

\bibitem{acerdet}
  E.~Richter-Was,
  arXiv:hep-ph/0207355.

\bibitem{csc_notes}
 G.~Aad {\it et al.}  [The ATLAS Collaboration],
  arXiv:0901.0512 [hep-ex].

\bibitem{ayres}
  for a similar conclusion see \eg   
  A.~Freitas and D.~Wyler,
  JHEP {\bf 0611}, 061 (2006).

\bibitem{fatjet_w}
 M.~H.~Seymour,
  Z.\ Phys.\  C {\bf 62}, 127 (1994);
 J.~M.~Butterworth, B.~E.~Cox and J.~R.~Forshaw,
  Phys.\ Rev.\  D {\bf 65}, 096014 (2002);
 W.~Skiba and D.~Tucker-Smith,
  Phys.\ Rev.\  D {\bf 75}, 115010 (2007);
 B.~Holdom,
  JHEP {\bf 0703}, 063 (2007).

\bibitem{fatjet_susy}
 J.~M.~Butterworth, J.~R.~Ellis and A.~R.~Raklev,
  JHEP {\bf 0705}, 033 (2007);
 J.~M.~Butterworth, J.~R.~Ellis, A.~R.~Raklev and G.~P.~Salam,
  arXiv:0906.0728 [hep-ph];
 C.~S.~Cowden, S.~T.~French, J.~A.~Frost and C.~G.~Lester [The ATLAS Collaboration],
  ATLAS-PHYS-PUB-2009-076, June~2009;
 G.~D.~Kribs, A.~Martin, T.~S.~Roy and M.~Spannowsky,
  arXiv:0912.4731 [hep-ph]
 and
  arXiv:1006.1656 [hep-ph].


\bibitem{fatjet_higgs}
 C.~R.~Chen, M.~M.~Nojiri and W.~Sreethawong,
  arXiv:1006.1151 [hep-ph];
 A.~Falkowski, D.~Krohn, J.~Shelton, A.~Thalapillil and L.~T.~Wang,
  arXiv:1006.1650 [hep-ph].

\bibitem{ca_algo}
 Y.~L.~Dokshitzer, G.~D.~Leder, S.~Moretti and B.~R.~Webber,
  JHEP {\bf 9708}, 001 (1997);
 M.~Wobisch and T.~Wengler,
  arXiv:hep-ph/9907280.

\bibitem{fastjet}
 M.~Cacciari and G.~P.~Salam,
  Phys.\ Lett.\  B {\bf 641}, 57 (2006);
 M.~Cacciari, G.~P.~Salam and G.~Soyez, \url{http://fastjet.fr}

\bibitem{mt2}
 C.~G.~Lester and D.~J.~Summers,
  Phys.\ Lett.\  B {\bf 463}, 99 (1999);
 A.~Barr, C.~Lester and P.~Stephens,
  J.\ Phys.\ G {\bf 29}, 2343 (2003).

\bibitem{tt_resonance}
 see \eg 
 U.~Baur and L.~H.~Orr,
  Phys.\ Rev.\  D {\bf 77}, 114001 (2008);
 P.~Fileviez Perez, R.~Gavin, T.~McElmurry and F.~Petriello,
  Phys.\ Rev.\  D {\bf 78}, 115017 (2008);
 Y.~Bai and Z.~Han,
  JHEP {\bf 0904}, 056 (2009).

\bibitem{top_tag}
 K.~Agashe, A.~Belyaev, T.~Krupovnickas, G.~Perez and J.~Virzi,
  Phys.\ Rev.\  D {\bf 77}, 015003 (2008);
  M.~Gerbush, T.~J.~Khoo, D.~J.~Phalen, A.~Pierce and D.~Tucker-Smith,
   Phys.\ Rev.\  D {\bf 77}, 095003 (2008);
 G.~Brooijmans,
  ATL-PHYS-CONF-2008-008 and ATL-COM-PHYS-2008-001, Feb.~2008
 J.~Thaler and L.~T.~Wang,
  JHEP {\bf 0807}, 092 (2008);
 L.~G.~Almeida, S.~J.~Lee, G.~Perez, G.~Sterman, I.~Sung and J.~Virzi,
  Phys.\ Rev.\  D {\bf 79}, 074017 (2009);
 L.~G.~Almeida, S.~J.~Lee, G.~Perez, I.~Sung and J.~Virzi,
  Phys.\ Rev.\  D {\bf 79}, 074012 (2009);
 D.~Krohn, J.~Thaler and L.~T.~Wang,
  arXiv:0903.0392 [hep-ph].

\bibitem{fatjet_wash}
 S.~D.~Ellis, C.~K.~Vermilion and J.~R.~Walsh,
  arXiv:0903.5081 [hep-ph]
 and
  arXiv:0912.0033 [hep-ph].

\bibitem{giacinto}
 ATLAS note,
 ATL-PHYS-PUB-2009-088.

\bibitem{lhc_ilc}
 J.~Hisano, K.~Kawagoe and M.~M.~Nojiri,
  Phys.\ Rev.\  D {\bf 68}, 035007 (2003);
 G.~Weiglein {\it et al.}  [LHC/LC Study Group],
  Phys.\ Rept.\  {\bf 426}, 47 (2006).

\bibitem{sfitter}
 R.~Lafaye, T.~Plehn, M.~Rauch and D.~Zerwas,
  Eur.\ Phys.\ J.\  C {\bf 54}, 617 (2008);
 P.~Bechtle, K.~Desch, M.~Uhlenbrock and P.~Wienemann,
  [arXiv:0907.2589 [hep-ph]].


\end{thebibliography}
\end{document}